\documentclass[11pt,a4paper]{article}
\usepackage{jheppub}
\usepackage{mathtools,accents,mleftright,slashed,orcidlink,cleveref,bbm,booktabs,microtype}
\title{Localisation with on-shell supersymmetry algebras via the Batalin--Vilkovisky formalism:\\Localisation as gauge fixing}
\author{Leron Borsten\textsuperscript{\orcidlink{0000-0001-9008-7725}},}
\author{Dimitri Kanakaris\textsuperscript{\orcidlink{0009-0001-7716-851X}},}
\author{and Hyungrok Kim\textsuperscript{\orcidlink{0000-0001-7909-4510}}}
\affiliation{Centre for Mathematics and Theoretical Physics, Department of Physics, Astronomy and Mathematics, University of Hertfordshire, Hatfield, Hertfordshire AL10 9AB, United Kingdom}
\emailAdd{l.borsten@herts.ac.uk}
\emailAdd{d.kanakaris-decavel@herts.ac.uk}
\emailAdd{h.kim2@herts.ac.uk}
\abstract{
The Batalin--Vilkovisky formalism provides a powerful technique to deal with gauge and global (super)symmetries that may only hold on shell.
We argue that, since global (super)symmetries and gauge symmetries appear on an equal footing in the Batalin--Vilkovisky formalism,
similarly localisation with respect to global (super)symmetries appears on an equal footing with gauge fixing of gauge symmetries; in general, when the gauge-fixing condition is not invariant under the global symmetries, localisation (with respect to a localising fermion) and gauge fixing (with respect to a gauge-fixing fermion) combine into a single operation.
Furthermore, this perspective enables supersymmetric localisation using only on-shell supermultiplets, dispensing with auxiliary fields, extending an insight first discovered by Losev and Lysov \cite{Losev:2023gsq}. We provide the first examples of on-shell localisation for quantum field theories (together with a companion paper by Arvanitakis~\cite{Arvanitakis:2025}).
}
\keywords{supersymmetric localisation, Batalin--Vilkovisky formalism, on-shell symmetry}
\hypersetup{
  pdfauthor={Leron Borsten, Dimitri Kanakaris Decavel, 金炯錄},
  pdftitle={Localisation with on‐shell supersymmetry algebras via the Batalin–Vilkovisky formalism: localisation as gauge fixing},
  pdfsubject={hep-th},
  pdflang={en-GB},
  pdfkeywords={supersymmetric localisation, Batalin–Vilkovisky formalism, on‐shell symmetry},
}

\newcommand{\frg}{\mathfrak{g}}
\newcommand{\frr}{\mathfrak{r}}
\newcommand{\frs}{\mathfrak{s}}
\newcommand{\frF}{\mathfrak{F}}
\newcommand{\frR}{\mathfrak{R}}
\newcommand{\frS}{\mathfrak{S}}

\newcommand{\bbN}{\mathbb{N}}
\newcommand{\bbZ}{\mathbb{Z}}
\newcommand{\bbR}{\mathbb{R}}
\newcommand{\bbC}{\mathbb{C}}
\newcommand{\idop}{\mathbbm{1}}
\newcommand{\cA}{\mathcal{A}}
\newcommand{\cB}{\mathcal{B}}
\newcommand{\cO}{\mathcal{O}}
\newcommand{\cL}{\mathcal{L}}
\newcommand{\cN}{\mathcal{N}}
\newcommand{\cG}{\mathcal{G}}
\newcommand{\cP}{\mathcal{P}}
\newcommand{\cQ}{\mathcal{Q}}

\newcommand{\sfb}{\mathsf{b}}
\newcommand{\sfi}{\mathsf{i}}
\newcommand{\sfp}{\mathsf{p}}
\newcommand{\sfr}{\mathsf{r}}
\newcommand{\sft}{\mathsf{t}}
\newcommand{\sfR}{\mathsf{R}}
\newcommand{\sfC}{\mathsf{C}}
\newcommand{\rme}{\mathrm{e}}
\newcommand{\rmi}{\mathrm{i}}
\newcommand{\sigmat}{\accentset{\sim}{\sigma}}
\newcommand{\betat}{\accentset{\sim}{\beta}}
\newcommand{\zetat}{\accentset{\sim}{\zeta}}
\newcommand{\lambdat}{\accentset{\sim}{\lambda}}

\newcommand{\Symm}{\mathsf{Symm}}
\newcommand{\symm}{\mathsf{symm}}
\newcommand{\osSymm}{\mathsf{S\overline{ymm}}}
\newcommand{\ossymm}{\mathsf{\overline{symm}}}
\renewcommand{\Gauge}{\mathsf{Gauge}}
\newcommand{\gauge}{\mathsf{gauge}}
\newcommand{\osGauge}{\mathsf{G\overline{auge}}}
\newcommand{\osgauge}{\mathsf{\overline{gauge}}}
\newcommand{\Global}{\mathsf{Global}}
\newcommand{\globall}{\mathsf{global}}
\newcommand{\Trivial}{\mathsf{Trivial}}
\newcommand{\trivial}{\mathsf{trivial}}
\newcommand{\crit}{\mathsf{crit}}
\newcommand{\BRST}{\mathsf{BRST}}
\newcommand{\BV}{\mathsf{BV}}
\newcommand{\even}{\mathsf{even}}
\newcommand{\odd}{\mathsf{odd}}
\newcommand{\id}{\mathsf{id}}
\newcommand{\gf}{\mathsf{gf}}
\newcommand{\BPS}{\mathsf{BPS}}
\newcommand{\loc}{\mathsf{loc}}
\newcommand{\gBRST}{{\mathsf{gBRST}}}
\newcommand{\CE}{{\mathsf{CE}}}
\newcommand{\gBV}{{\mathsf{gBV}}}
\newcommand{\off}{\mathsf{off}}
\newcommand{\on}{\mathsf{on}}
\newcommand{\free}{\mathsf{free}}
\newcommand{\old}{\mathsf{old}}
\newcommand{\new}{\mathsf{new}}

\newcommand{\ym}{\textsc{ym}}
\newcommand{\SYM}{\mathsf{SYM}}
\newcommand{\lin}{\mathsf{lin}}
\newcommand{\Diffeo}{\operatorname{{Diffeo}}}
\newcommand{\Lie}{\operatorname{{Lie}}}
\renewcommand{\Im}{\operatorname{{Im}}}
\newcommand{\gh}{\operatorname{\mathsf{gh}}}
\newcommand{\rank}{\operatorname{{rank}}}
\newcommand{\Hess}{\operatorname{{Hess}}}
\renewcommand{\div}{\operatorname{{div}}}
\newcommand{\Ber}{\operatorname{{Ber}}}
\newcommand{\Det}{\operatorname{{Det}}}

\newcommand{\ad}{\operatorname{{ad}}}
\newcommand{\Ad}{\operatorname{{Ad}}}
\newcommand{\Order}{O}
\newcommand{\sign}{\operatorname{{sign}}}

\newcommand{\tr}{\operatorname{{tr}}}
\newcommand{\tT}{\mathrm{T}}
\newcommand{\tN}{\mathrm{N}}
\newcommand{\into}{\hookrightarrow}
\newcommand{\onto}{\twoheadrightarrow}
\newcommand{\Phia}{\Phi^+}
\newcommand{\phia}{\phi^+}
\newcommand{\Xia}{\Xi^+}
\newcommand{\xia}{\xi^+}
\newcommand{\asta}{\#}
\newcommand{\+}{\times}
\newcommand{\del}{\partial}
\newcommand{\pder}[2]{\frac{\del #1}{\del #2}}
\newcommand{\tpder}[2]{\tfrac{\del #1}{\del #2}}
\newcommand{\fder}[2]{\frac{\delta #1}{\delta #2}}
\newcommand{\tfder}[2]{\tfrac{\delta #1}{\delta #2}}
\newcommand{\deltar}{\accentset{\leftarrow}{\delta}}
\newcommand{\deltal}{\accentset{\rightarrow}{\delta}}
\newcommand{\fderl}[2]{\frac{\deltal #1}{\delta #2}}
\newcommand{\fderr}[2]{\frac{\deltar #1}{\delta #2}}
\newcommand{\tfderl}[2]{\tfrac{\deltal #1}{\delta #2}}
\newcommand{\tfderr}[2]{\tfrac{\deltar #1}{\delta #2}}
\newcommand{\pderr}[2]{\frac{\accentset{\leftarrow}{\del} #1}{\del #2}}
\newcommand{\pderl}[2]{\frac{\accentset{\rightarrow}{\del} #1}{\del #2}}
\newcommand{\tder}[2]{\tfrac{\extd #1}{\extd #2}}
\newcommand{\der}[2]{\frac{\extd #1}{\extd #2}}
\newcommand{\normeq}{\trianglelefteq}
\newcommand{\morneq}{\trianglerighteq}
\newcommand{\extd}{\mathrm{d}}
\newcommand{\extD}{\mathrm{D}}
\newcommand{\EV}[1]{\left\langle #1\right\rangle}
\newcommand{\overeq}[1]{\overset{\text{#1}}{=}}
\newcommand{\inp}{\mathbin{\lrcorner}}
\makeatletter
\newcommand{\hathat}[1]{% 
\begingroup%
  \let\macc@kerna\z@%
  \let\macc@kernb\z@%
  \let\macc@nucleus\@empty%
  \hat{\raisebox{.3ex}{\vphantom{\ensuremath{#1}}}\smash{\hat{#1}}}%
\endgroup%
}
\makeatother

\newcommand{\zetab}{\bar{\zeta}}
\newcommand{\tH}{{\text{H}}}
\newcommand{\bos}{\mathsf{bos}}
\newcommand{\fer}{\mathsf{fer}}

\begin{document}
\maketitle

\section{Introduction}\label{sec:intro}
Localisation techniques \cite{Pestun:2007rz,Pestun:2016zxk,Pestun:2016jze,Karki:1993bw,Szabo:1996md} provide a powerful tool to compute expectation values and correlation functions of operators that are invariant  under certain global symmetries, a prime example being Bogomol'nyi–Prasad–Sommerfeld (BPS) states in supersymmetric quantum field theory. 
In this technique, the path integral over all possible field configurations is reduced to one over a smaller space of field configurations invariant under a subset of the symmetries, which can often be evaluated explicitly.
On the other hand, gauge fixing provides a method to compute expectation values and correlation functions of operators that are invariant under gauge symmetries.
In this technique, the path integral over all possible field configurations is reduced to one over a smaller space of field configurations satisfying a gauge-fixing condition. The two techniques of localisation and gauge fixing appear intriguingly similar.

In this paper, we make precise the connection between localisation and gauge fixing: in the Batalin--Vilkovisky (BV) formalism of quantum field theory \cite{Batalin:1977pb,Batalin:1981jr,Batalin:1983ggl,Batalin:1984ss,Batalin:1985qj}, the two are but different special cases of a single operation of restricting the path integral to a submanifold by means of symmetries.
Just as one performs gauge fixing with a gauge-fixing fermion, localisation can be phrased in terms of an analogous \emph{localising fermion}.
Furthermore, when one gauge-fixes with a gauge-fixing condition that does not respect some of the global symmetries of the theory, the two procedures of gauge-fixing and localisation entwine into a single procedure with respect to a more general fermion \(\Psi\) \eqref{gauge fixing localising fermion} that combines and subsumes the gauge-fixing and localising fermions together with some mixing terms. Besides the conceptual clarity, since the BV formalism was purpose built to handle \emph{open} gauge algebras (i.e.~that close only on-shell), this unification provides a framework for localisation with respect to open global (super)symmetries, as will be demonstrated in several examples.

\paragraph{An illustrative primer:   supersymmetric Maxwell theory.} The localisation/Batalin--Vilkovisky parallel is sufficiently   manifest to  be sketched  for the uninitiated reader using  (Euclidean)  supersymmetric Maxwell theory\footnote{For which neither localisation nor the BV formalism are needed, since the partition function may be evaluated directly. Nonetheless, even this simple model has its surprises. See, for example, \cite{Donnelly:2016mlc,Borsten:2021pte,Borsten:2025phf}.}, with $\mathrm U(1)$ gauge potential $A_\mu$ and photino $\lambda_\alpha$\footnote{There may be more photini and scalar fields, but they are not important to the key concepts, so let us not unnecessarily clutter the parallel with them.}.  Let $S^{\textsf{free}}_{\SYM}$ be the  supersymmetric Maxwell action, invariant under  a \emph{global} supersymmetry 
\begin{equation}
\delta_\varepsilon A_\mu = \varepsilon_\alpha \gamma^{\alpha\beta}_\mu \lambda_\beta,\qquad  \delta_\varepsilon \lambda_\beta = \slashed{F}_\alpha{}^\beta\varepsilon_\beta
\end{equation}
and a \emph{local} gauge symmetry
\begin{equation}
\delta_\theta A_\mu = \partial_\mu \theta,\qquad  \delta_\theta \lambda_\alpha = \mathrm i \theta \lambda_\alpha\,.
\end{equation}
The supersymmetry corresponds to a  supercharge $\cQ_\alpha$  that is nilpotent, $[\cQ_\alpha, \cQ_\beta]=0$, such that $\delta_\varepsilon = \varepsilon^\alpha\cQ_\alpha$. Similarly, by promoting the local gauge parameter $\theta$ to a ghost field $c$, the gauge symmetry corresponds to a BV  charge $Q_\BV$  that is nilpotent, $Q_\BV^2=0$, such that  $\delta_c = Q_\BV$ on the fields.

On the localisation side, with appropriate assumptions the partition function is invariant under deformations by a $\cQ$-exact term 
\begin{equation}
S^{\textsf{free}}_{\SYM}\mapsto S^{\textsf{free}}_{\SYM} + t \cQ \Psi_{\loc}\,,
\end{equation}
where $t\in \bbR$ and $\Psi_{\loc}$ is the localising fermion. Taking the $t\to\infty$ limit (again with appropriate assumptions) localises the partition function to an integral over supersymmetry orbits for BPS states,  $\cQ \cO_\BPS=0$. 

On the Batalin--Vilkovisky side, with appropriate assumptions the partition function is invariant under deformations by a $Q_\BV$-exact term 
\begin{equation}
S^{\textsf{free}}_{\SYM}\mapsto S^{\textsf{free}}_{\SYM} + Q_\BV \Psi_{\gf}\,,
\end{equation}
where  $\Psi_{\gf}$ is the gauge-fixing  fermion. Here,  (with appropriate assumptions) the partition function localises  to an integral over gauge orbits for gauge-invariant states,  $Q_{\BV} \cO=0$. 

While the above wording  makes the similarities manifest, differences remain; these must be harmonised to fulfil the proposed unification. The first key disanalogy (which cannot, and need not, be resolved), is that while the localising deformation  $t \cQ \Psi_{\loc}$ is optional, the gauge-fixing deformation $Q_\BV \Psi_{\gf}$ is obligatory. The second difference (which can, and must, be reconciled) is that $\cQ$ has fermion parity $\odd\in \bbZ_2$ and ghost number $0\in \bbZ$, while $Q_\BV$ has fermion parity $\even \in \bbZ_2$ and ghost number $+1 \in \bbZ$. This is straightforwardly resolved by noting that we mapped the ghost-number-zero local gauge parameter $\theta$ to the local ghost $c$, so that we should similarly map the ghost-number-zero global supersymmetry parameter $\varepsilon$ to a \emph{global} ghost $\varepsilon$ (which we donate with the same symbol). Letting $Q=\varepsilon^\alpha\cQ_\alpha$, the supercharge $Q$ and the BV charge $Q_\BV$ are then placed on the same footing. This is a well-known (to BV aficionados, at least) manoeuvre \cite{Brandt:1996uv,Brandt:1996nr,Brandt:1997cz, Jurco:2018sby}. Finally, $\Psi_{\loc}$ has fermion number $1$ and ghost number $0$, while $\Psi_\gf$ has fermion number $0$ and ghost number $-1$. To harmonise, first note that, since $A_\mu, \lambda, c$ carry non-negative ghost number, to construct $\Psi_\gf$  with ghost number $-1$, one must introduce a trivial\footnote{Trivial in the sense that they have vanishing $Q_\BV$ cohomology and so leave the physics invariant.} pair $(\bar c, b)$, i.e.~the  antighost $\bar c$ with ghost number $-1$ and the Nakanishi--Lautrup field $b$. Then $\Psi_\gf\sim \bar c G$, where $G$ is the gauge-fixing function (typically $G\sim  \partial^\mu A_\mu + t b$). So, following the parallel, one ought to  introduce a trivial pair $(\bar \sigma, \beta)$, where $\bar \sigma$ is the antighost corresponding to the global ghost $\varepsilon$. Then $\Psi_\loc\sim \bar \sigma V$, where $V$ is the localising  potential (typically $V\sim \lambda (\cQ \lambda)^\dagger+t\beta$). At this stage, it becomes clear that localisation and gauge-fixing are, schematically at least, rendered equivalent via the Batalin--Vilkovisky formalism as summarised here:
\\

 \begin{center}  
   \begin{tabular}{@{} ll @{}} 
\toprule     
      Batalin--Vilkovisky gauge fixing    & Supersymmetric localisation \\
      \midrule
      Gauge symmetry $Q_\BV$      & Global symmetry $\cQ$ \\
       Gauge ghost $c$         & Global ghost $\varepsilon$      \\
             Gauge trivial pair  $(\bar c, b )$       &  Global trivial pair  $(\bar \sigma, \beta )$    \\
      Gauge-fixing fermion $\Psi_\gf$       &  Localising fermion $\Psi_\loc$     \\
      Gauge-fixing action $S+Q_\BV\Psi_\gf $       &  Localising action $S+t^2Q\Psi_\loc $   \\
      \bottomrule
   \end{tabular}
\end{center}
This is not unexpected; aside from the optional/obligatory distinction, the Batalin--Vilkovisky apparatus itself cannot know what symmetries to incorporate or not. What is sauce for the local goose is sauce for the global gander.

Of course, there are other approaches to this picture, e.g.~\cite{Losev:2023gsq,Lysov:2024lge, Arvanitakis:2025}. Indeed, here we also consider an alternative, and computationally convenient, construction of the localisation fermion.  Having  localised\footnote{In the sense of ring theory.} the ghost ring to include $1/\varepsilon$, which carries  ghost number $-1$, we can directly construct the localising fermion without introducing trivial pairs. If   $\Psi_\loc$ is a desirable ghost number $0$ localising fermion (in the sense that it will localise the path integral), then $\Psi_\loc/\varepsilon$ is a computationally convenient choice of ghost number $-1$ localisation fermion in the Batalin--Vilkovisky approach\footnote{Assuming the BV action is independent of the antifield $\varepsilon^+$, which is generically true.}. 

\paragraph{Localisation for on-shell (super)algebras.} Although the conceptual unification described in the proceeding paragraph is appealing, one should ask what it is good for. The original motivation underpinning the BV formalism, i.e.~gauge algebras that only close on-shell, immediately suggests one such application. 
Accordingly,  we explain how to perform localisation with on-shell realisations of symmetries via the BV formalism, extending the analysis of \cite{Losev:2023gsq,Lysov:2024lge}. Traditional discussions of localisation require auxiliary fields to realise supersymmetry algebras off shell, which leads to complications; moreover, beyond eight supercharges it is often impossible to realise the supersymmetry algebra off shell without an infinite number of auxiliary fields (as in the pure-spinor formalism \cite{Eager:2021wpi}). However, the Batalin--Vilkovisky formalism treats on-shell and off-shell realisations uniformly using the powerful techniques of homological algebra, so that localisation can be performed for on-shell realisations of symmetries just as easily.
On-shell localisation was first discussed in \cite{Losev:2023gsq,Lysov:2024lge} in the context of finite-dimensional models; this paper presents the first computations of on-shell localisation for supersymmetric field theories to our knowledge (together with a companion paper by Arvanitakis~\cite{Arvanitakis:2025}), namely for the \(d=1\), \(\mathcal N=2\) superparticle as well as the \(d=3\), \(\mathcal N=2\) supersymmetric Yang--Mills theory on a Seifert manifold.

\paragraph{Related work.}
The idea of applying localisation techniques to on-shell symmetries using the Batalin--Vilkovisky formalism is not new. For ordinary (bosonic) symmetries (i.e.\ equivariant localisation), this appeared in early work by Nersessian \cite{Nersessian:1993eq,Nersessian:1993me} and Kalkman \cite{Kalkman:1993zp}. For supersymmetries, this first appeared recently in work by Losev and Lysov \cite{Losev:2023gsq,Lysov:2024lge}. In all these papers, however, the examples considered are finite-dimensional systems rather than field theories.
The recent work by Cattaneo and Jiang \cite{Cattaneo:2025wdw} discusses equivariant localisation in the context of topological field theories; its concluding paragraph also mentions a connection to gauge fixing in the Batalin--Vilkovisky formalism.
Finally, we mention the  related work by Arvanitakis~\cite{Arvanitakis:2025}, which arose from a collaboration with the present authors and which is being published on the arXiv at the same time as this paper (cf.\ the Acknowledgements).

\paragraph{Future directions.} In the present contribution, we restricted ourselves to the minimal formalism required to apply the BV apparatus to  examples that illustrate the procedure concretely, but concisely. To go beyond this, in a subsequent paper to appear, we will (1) further develop the general formalism to include trivial pairs (a.k.a.\ antighosts and Nakanishi--Lautrup fields) for localisation, (2) apply this to various $d=4$ supersymmetric quantum field theories. Regarding (1), in \cref{sec:superparticle} and \cref{sec:seifert}  we give concrete examples of the trivial pair construction for the superparticle and $d=3, \mathcal{N}=2$ supersymmetric Yang--Mills theory, respectively. 

We also restrict ourselves for simplicity to ordinary symmetries, i.e.\ those described by a Lie algebra or superalgebra. A powerful feature of the Batalin--Vilkovisky formalism is that it naturally describes higher symmetries (both gauge and global).
Higher gauge symmetries appear in higher gauge theory \cite{Borsten:2024gox}, and twisted theories exhibit higher spacetime symmetry \cite{Jonsson:2024uyr,Hahner:2025nmk} and higher correlation functions \cite{Alfonsi:2025kmj}. We expect the methods of this paper to generalise straightforwardly to the case of such higher symmetries.

On many topologies, one must couple to background supergravity fields to preserve supersymmetry and localise \cite{Festuccia:2011ws,Dumitrescu:2016ltq}.
Using the methods of this paper, one should still be able to dispense with the auxiliary fields,  working  with an on-shell supermultiplet \cite{Borsten:2025hrn}. 

Finally, it is often possible to introduce a `fake' or `evanescent' supersymmetry to non-supersymmetric theories without changing the physics \cite{Jurco:2018sby,Arvanitakis:2024dbu}; for instance, Chern--Simons theory may be evanescently supersymmetrised by adjoining auxiliary fields \cite{Lee:1990it}, which is equivalent to the non-supersymmetric theory even at the quantum level \cite{Fan:2018wya}. This evanescent supersymmetry may still be used to perform localisation computations \cite{Beasley:2005vf,Kapustin:2009kz,Arvanitakis:2024dbu}. Often realising evanescent supersymmetry off shell requires adding auxiliary fields (as for Chern--Simons theory), and we expect that our methods may be used to simplify such computations. In particular, the so-called Manin class of theories \cite{Arvanitakis:2024dbu,Kanakaris:2023kfq,Borsten:2024pfz,Borsten:2024alh,Arvanitakis:2025nyy,Borsten:2025pvq} that appear in various contexts may be amenable to such computations.

\paragraph{Organisation of this paper.}
This paper is organised as follows. After a brief review of the Batalin--Vilkovisky formalism in \cref{sec:bv review}, we reformulate the usual localisation procedure for off-shell symmetries using the notion of a localising fermion (analogous to the gauge-fixing fermion) in \cref{sec:off shell localisation}. This sets the stage for \cref{sec:on shell localisation}, which generalises the usual localisation argument to also apply to symmetries that only hold on shell. Finally, \cref{sec:superparticle} and \cref{sec:seifert} provide examples of the on-shell localisation procedure for the \(\mathcal N=2\) superparticle in a superpotential and for three-dimensional \(\mathcal N=2\) supersymmetric Yang--Mills theory on a Seifert manifold, respectively.

\section{Lightning review of the Batalin--Vilkovisky formalism and symmetries}\label{sec:bv review}
We first briefly review our weapon of choice, the Batalin--Vilkovisky formalism and the associated language of graded geometry, setting up notation for subsequent sections. For more detailed reviews, we refer to \cite{Gomis:1994he,Henneaux:1994lbw,Fiorenza:2004sg,Fuster:2005eg,Qiu:2011qr,Barnich:2018gdh}. In what follows, the \(\bbZ\times\bbZ_2\)-valued bidegree of a coordinate function \(\phi^i\) on a graded manifold will be denoted \(|\phi^i|\), and the total Grassmann parity of a coordinate function \(\phi^i\) with \(|\phi^i|=(a,b)\) is
\(\|\phi^i\|\coloneqq(a\bmod 2)+b\in\bbZ_2\), where \(i\) is a DeWitt index that includes spacetime position as well as any Lorentz and discrete indices; derivatives with respect to it are thus functional derivatives.

The Batalin--Vilkovisky formalism starts with the infinite-dimensional space of all spacetime configurations (or histories) of physical fields as an infinite-dimensional manifold \(\frF\) graded by \(\bbZ_2\) (i.e.\ a supermanifold).
The action \(S\colon\frF\to\bbR\) is a real-valued function on \(\frF\), whose critical surface \(\frF_\crit \coloneqq \{x\in\frF|\mathrm dS|_x = 0\}\) is the submanifold of solutions to the equations of motion.
This \(\bbZ_2\)-graded manifold is acted upon by a group \(\Gauge\) of gauge symmetries. Under suitable assumptions (see \cite{Henneaux:1994lbw}), using homological perturbation theory one can resolve the possibly singular quotient \(\frF/\Gauge\) into a \(\bbZ\times\bbZ_2\)-graded manifold \(\frF_\BV \coloneqq\tT^*[-1]\frF_\BRST\) (which is the shifted cotangent space of the BRST manifold \(\frF_\BRST\)) together with a Batalin--Vilkovisky differential \(Q_\BV\) of bidegree \(|Q_\BV| = (+1,\even)\) on \(\frF_\BV\), which is a resolution in the homological sense:
\begin{equation}\label{eq:funconphase}
	H^0(Q_\BV) \eqqcolon C^\infty\mleft(\frF_\crit\big/\Gauge\mright)\,.
\end{equation}
One should formally regard $C^\infty\mleft(\frF_\crit\big/\Gauge\mright)$ as the ring of functionals on the phase space \(\Omega = \frF_\crit/\Gauge\). Note, such quotients are generically singular and should be defined in terms of the corresponding  action Lie algebroid (or derived quotient) as in \eqref{eq:funconphase}.\footnote{Moreover, the spaces in consideration will be infinite-dimensional, which requires regularisation for the  BV Laplacian below (and the BV symplectic form in the case of non-compact spacetimes). However, for all our examples this issue does not arise, since the classical BV action will solve the quantum master equation.}

The Batalin--Vilkovisky manifold \(\frF_\BV\) naturally carries a symplectic form \(\omega_\BV\) of degree \((-1,\even)\), an antibracket \((-,-)_\BV\), and degree-\((+1,\even)\) BV Laplacian \(\Delta_\BV\) given by
\begin{subequations}
\label{BV symplectic form etc}
\begin{align}
	\omega_\BV &\coloneqq (-)^{\|\Phi^I\|}\delta\Phi^I\wedge\delta\Phia_I\,,
	\\
	(F,G)_\BV &\coloneqq F\Big(\fderr{}{\Phia_I}\fderl{}{\Phi^I} - \fderr{}{\Phi^I}\fderl{}{\Phia_I}\Big)G\,,
	\\
	\Delta_\BV &\coloneqq (-)^{\|\Phi^I\|}\fder{}{\Phi^I}\fder{}{\Phia_I}\,,
\end{align}
\end{subequations}
where the index \(I\) is a tangent index for \(\frF_\BRST\), so that \(\frF_\BRST\) has local coordinates \(\Phi^I\) and \(\frF_\BV=\tT^*[-1]\frF_\BRST\) has local coordinates \((\Phi^I,\Phia_I)\), and where \(F,G\in C^\infty_\bullet(\frF_\BV)\).\footnote{We take the functionals \(C_\bullet^\infty(\frF_\BV) = \bigoplus_{(k,p)\in\bbZ\times\bbZ_2}C_{(k,p)}^\infty(\frF_\BV)\) to be smooth in fields of vanishing ghost number and polynomial in fields of non-vanishing ghost number. By \(C^\infty(\frF_\BV)\) we will mean the superfunctions on the underlying supermanifold, which can be smooth in all variables.}
The Batalin--Vilkovisky differential \(Q_\BV = (S_\BV,-)_\BV\) is then the Hamiltonian vector field of the degree-\((0,\even)\) \emph{Batalin--Vilkovisky action} \(S_\BV \in C^\infty_{(0,\even)}(\frF_\BV)\), which satisfies the \emph{classical master equation}
\begin{equation}
	Q_\BV^2 = 0\,,
\end{equation}
or equivalently
\begin{equation}
	(S_\BV,S_\BV)_\BV = 0\,.
\end{equation}
The classical master equation together with suitable boundary conditions
uniquely determine the Batalin--Vilkovisky action up to canonical transformations and inclusion of trivial pairs.

\subsection{Gauge fixing as restriction to a Lagrangian submanifold}
In the Batalin--Vilkovisky formalism, gauge fixing corresponds to restriction of the path integral to a Lagrangian submanifold of \(\frF_\BV\) (paralleling the restriction to the symmetry-fixed submanifold in localisation) in the following sense. The upshot of the BV procedure is that we have replaced the gauge structure of the theory by a global BV symmetry \(Q_\BV\), as well as new gauge symmetries
\begin{align}
	R_I 
	&= \big(\fder{S_\BV}{\Phi^I},-\big)_\BV\,,
	&
	R_+^I 
	&= \big(\fder{S_\BV}{\Phia_I},-\big)_\BV\,.
\end{align}
Under reasonable assumptions \cite{Gomis:1994he}, this yields enough gauge symmetry to kill off half of the field content, so that gauge fixing corresponds to restricting the action to a Lagrangian submanifold \(\iota_\gf\colon \cL_\gf\into\frF_\BV\).
Such Lagrangian submanifolds are not unique;
a class of such Lagrangian submanifolds are obtained through the choice of a \emph{gauge-fixing fermion}, which is a function \(\Psi_\gf\) on \(\frF_\BRST\) of degree \((-1,\even)\). This then determines the Lagrangian submanifold
\begin{equation}
\begin{aligned}
	\iota_\gf\colon \frF_\BRST & \into \frF_\BV\\
	\Phi^I &\mapsto \left(\Phi^I,\Phia_I = \fder{\Psi_\gf}{\Phi^I}\right)\,,
\end{aligned}
\end{equation}
as the graph \(\mathsf{Graph}(\extd\Psi_\gf)\into\frF_\BV \cong \tT^\ast[-1]\frF_\BRST\) of the gradient of the gauge-fixing fermion.

\subsection{Quantisation}
To quantise the theory, one introduces the BV path integral measure
\begin{equation}
	\extD\mu_\BV(\hbar)
	=
	\underbrace{\extD\Phi\,\extD\Phia}_{\eqcolon \extD\mu_\BV} ~ \exp\mleft(-S_\BV^1 - \hbar S_\BV^2 - \hbar^2 S_\BV^3 - \cdots\mright)\,,
\end{equation}
where the \(S_\BV^g\) is the order \(\Order(\hbar^g)\) counterterms to be added to the action. We require that this measure be compatible with the BV differential \(Q_\BV\) in the sense that it be divergenceless, which is equivalent to the \emph{quantum master equation},
\begin{subequations}
\label{quantum master equation}
\begin{align}
	\Big.0 &= \div_{\extD\mu_\BV(\hbar)}Q_\BV ~~~~~~\&~~~~~~ 0 = Q_\BV^2
	\\
	\Big.\Leftrightarrow 0 &= \Delta_\BV e^{-\frac{1}{\hbar}S_\BV}
	\\
	\Big.\Leftrightarrow 0  &= \hbar\Delta_\BV S_\BV - \tfrac12(S_\BV,S_\BV)_\BV
\end{align}
\end{subequations}
where \(S_\BV =\sum_g \hbar^g S_\BV^g \in C_\bullet^\infty(\frF_\BV)[\![\hbar]\!]\) is the quantum corrected BV action.\footnote{Here, we take \(C_\bullet^\infty(\frF_\BV)[\![\hbar]\!]\)  the ring of formal power series  in \(\hbar\) valued in \(C^\infty_\bullet(\frF_\BV)\).} Similarly, we define quantum observables \(\cO\in C^\infty(\frF_\BV)[\![\hbar]\!]\) as satisfying
\begin{align}
\label{quantum observable}
	\Delta_\BV\Big\{\cO\mathrm e^{-\frac{1}{\hbar}S_\BV}\Big\} &= 0
	&
	&\Leftrightarrow
	&
	\hbar\Delta_\BV\cO - (S_\BV,\cO)_\BV &= 0\,.
\end{align}
Henceforth, we will set \(\hbar = 1\).

The measure \(\extD\mu_\BV = \extD\Phi\,\extD\Phia\in\Gamma(\Ber(\tT^\ast\frF_\BV)^\times)\) is not well defined on the Lagrangian submanifold \(\cL_\gf\into\frF_\BV\).
However, using the symplectic structure, one can canonically map densities to half-densities \(\extD\mu_\BV\mapsto\sqrt{\extD\mu_\BV} \in \Gamma(\sqrt{\Ber}(\tT^\ast\frF_\BV)^\times)\), which pull back to densities on the Lagrangian submanifold \(\cL_\gf\) as \(\iota_\gf^*\colon\Gamma(\sqrt{\Ber}(\tT^\ast\frF_\BV)^\times) \to \Gamma(\Ber(\tT^\ast\cL_\gf)^\times)\) \cite{Kudaverdian:2000xn}. The expectation value of a quantum observable \(\cO \in C^\infty(\frF_\BV)\) is thus given by
\begin{equation}
\label{expectation value}
	\EV{\cO}
	=
	\int_{\cL_\gf}\sqrt{\extD\mu_\BV} ~ \cO\mathrm e^{-S_\BV}\,.
\end{equation}
In particular, \(\iota[\Psi_\gf]^*\sqrt{\extD\Phi\,\extD\Phia} = \extD\Phi \in \Gamma(\Ber(\tT^\ast\frF_\BRST)^\times)\). Furthermore, equations \eqref{quantum master equation} and \eqref{quantum observable} then imply that \eqref{expectation value} is invariant under small perturbations \(\delta\Psi_\gf\) of the gauge-fixing fermion.

\subsection{Blurring the distinction between global and gauge symmetries}
\label{ssec:Blurring the distinction between global and gauge symmetries}
Let us further discuss global and gauge symmetries for this theory. (We assume that anomalies are absent wherever this is relevant.) We denote the group of symmetries of the theory by \(\Symm \leq \Diffeo(\frF)\), where $\Symm 
	\coloneqq \big\{\varphi\in\Diffeo(\frF)\,\big|\,S\circ\varphi = S\big\}$. We take the group of gauge symmetries to be a \emph{normal} subgroup of the group of symmetries, denoted by \(\Gauge \trianglelefteq \Symm\). 
The other important normal subgroup of $\Symm$ consists of the trivial symmetries \(\Trivial \trianglelefteq \Symm\), which are those symmetries that pull back to the identity map on the critical surface \(\varphi\circ \iota_\crit = \operatorname{id}_{\frF_\crit}\), where \(\iota_\crit\colon\frF_\crit\into\frF\), cf.\ \cite{MartinezAlonso:1979fej}, \cite[p.~377]{zbMATH00478435}, \cite[§3.1.5]{Henneaux:1994lbw}. We include, by definition, the  trivial symmetries in  the gauge symmetries, \(\Trivial\trianglelefteq\Gauge\), which will not affect the physics in any way. We can then quotient the (gauge) symmetries by the trivial symmetries to obtain the group of honest (gauge) symmetries \(\osSymm\) (resp.\ \(\osGauge\)).
The group of global symmetries \(\Global\) is then defined to be the quotient of the symmetries by the gauge symmetries. This definition  accommodates  global symmetries that only close up to (honest) gauge symmetries and trivial symmetries, as is often the case for supersymmetry. In summary, 
\begin{subequations}
\begin{align}
	\Symm 
	&= \big\{\varphi\in\Diffeo(\frF)\,\big|\,S\circ\varphi = S\big\}
	\,,
	\\
	\Gauge &\trianglelefteq \Symm
	\,,
	\\
	\Trivial &= \big\{\varphi\in\Symm\,\big|\,\varphi\circ\iota_\crit = \id\big\} \trianglelefteq \Gauge\,.
\end{align}
\end{subequations}
We may pass to the associated Lie algebras:
\begin{subequations}
\begin{align}
	\symm \coloneqq \Lie(\Symm)
	&= \big\{r\in\Gamma(\tT\frF)\,\big|\,rS = 0\big\}\,,
	\\
	\gauge \coloneqq \Lie(\Gauge)
	&\trianglelefteq \symm\,,
	\\
	\trivial \coloneqq \Lie(\Trivial)
	&= \big\{\mu\in\symm\,\big|\,\iota_\crit^*\mu = 0\big\}\,,
\end{align}
\end{subequations}
where \(\trianglelefteq\) denotes normal subgroups or ideals for groups or Lie algebras, respectively, and \(\Lie\) denotes the Lie functor. Finally, we quotient these groups (resp.\ Lie algebras) to give
\begin{subequations}
\begin{align}
	\osSymm &\coloneqq \Symm\big/\Trivial\,,
	&
	\ossymm &\coloneqq \Lie(\osSymm)
	= \symm\big/\trivial\,,
	\\
	\osGauge &\coloneqq \Gauge\big/\Trivial\,,
	&
	\osgauge &\coloneqq \Lie(\osGauge)
	= \gauge\big/\trivial\,,
	\\
	\label{global definition}
	\Global &\coloneqq \Symm\big/\Gauge\,,
	&
	\globall &\coloneqq \Lie(\Global)
	= \symm\big/\gauge\,.
\end{align}
\end{subequations}
These groups of symmetries are generally very large: they include not only symmetry transformations, but also symmetry transformations with parameters that are explicitly field-dependent. One way this manifests itself is for global symmetries in gauge theory, where generically a global symmetry and a gauge symmetry commute into a field-dependent global symmetry.

The phase space \(\Omega\) of the theory is given by
\begin{equation}
	\Omega 
	\coloneqq \frF_\crit\big/\Gauge 
	= \frF_\crit\big/\osGauge\,.
\end{equation}
(This quotient is well-defined since symmetries map solutions to the field equations to other solutions.)
The second follows since \(\Trivial\) acts trivially on \(\frF_\crit\).
The classical observables of the theory are then given by the function ring \(C^\infty(\Omega)\) on the phase space. This definition of phase space is identical to that of the Hamiltonian formalism.

\section{Localisation for off-shell algebras as gauge fixing}\label{sec:off shell localisation}
Let us now review the localisation arguments when the supersymmetry algebra closes off-shell, which is the case that is considered in most of the contemporary literature \cite{Pestun:2016jze}.

\subsection{Assumptions about symmetry structure} 

In this context, we make three simplifying assumptions about the gauge structure:
\begin{itemize}
	\item[1.] \emph{Group symmetries:} The infinitesimal gauge symmetries \(\osgauge = \Gamma(\frg\times\frF_\crit)\) can be identified with the sections of an \emph{action} Lie algebroid \(\frg\times\frF_\crit\Rightarrow\frF_\crit\).  The anchor map of this Lie algebroid is induced by the gauge symmetry \(C^\infty(\frF)\)-module morphism \(R\colon  \frg\to\gauge\).
	Similarly, the finite gauge transformations \(\osGauge = \Gamma(\cG\times\frF_\crit)\) can be identified with the bisections of an \emph{action} groupoid \(\cG\times\frF_\crit \rightrightarrows\frF_\crit\). We refer to \(\cG\) as the \emph{gauge group} and \(\frg = \Lie(\cG)\) as the \emph{gauge algebra}. The fact that these consist of sections (resp.\ bisections) precisely encapsulates the fact that symmetry transformations may be field-dependent.
	\item[2.] \emph{Off-shell closure:} The representation of the gauge group \(\cG\) can be extended to all of history space \(\cG\times\frF\rightrightarrows\frF\), such that \(R:\frg\to\gauge\) defines the anchor map of the corresponding action Lie algebroid \(\frg\times\frF\Rightarrow\frF\).
	\item[3.] \emph{Irreducible gauge structure:} There are no higher gauge symmetries.
\end{itemize}
Furthermore, we will make similar simplifying assumptions about the global symmetries:
\begin{itemize}
	\item[1.] \emph{Group symmetries:} Infinitesimal global symmetries \(\globall = \Gamma(\frs\times\frF_\crit/\cG)\) are identified with the sections of an action Lie algebroid \(\frs\times\frF_\crit/\cG\Rightarrow\frF_\crit/\cG\). Similarly, finite global symmetries \(\Global = \Gamma(\frS\times\frF_\crit/\cG)\) are identified with bisections of an action groupoid \(\frS\times\frF_\crit/\cG\rightrightarrows\frF_\crit/\cG\).
	We refer to \(\frS\) as the \emph{global symmetry group} and \(\frs = \Lie(\frS)\) as the \emph{global symmetry algebra}.
	\item[2.] \emph{Off-shell closure:} The representation of the global symmetry group \(\frS\) can be extended to the coset space \(\frF/\cG\) as \(\frS\times\frF/\cG\rightrightarrows\frF/\cG\), and the respective global symmetry algebra \(\frs\) to an action Lie algebroid \(\frs\times\frF/\cG\Rightarrow\frF/\cG\).
	\item[3.] \emph{Irreducible global symmetry structure:} 
	\(\frS\) is represented faithfully on \(\frF/\cG\).
\end{itemize}
Even when both the gauge and global symmetries have an underlying group structure, the group of all symmetries need not have an underlying group structure; for example, in supersymmetric Yang--Mills theory, the global symmetries generally only close up to field-dependent gauge transformations.

\paragraph{(Super)symmetric observables.} 

Localisation computes correlation functions of observables that are invariant under a subgroup \(\frR\leq\frS\) of the global symmetries. Considering that \(\frR\) acts on \(\Omega = \frF_\crit/\cG\), these observables are given by
\begin{equation}
	C^\infty(\Omega/\frR)
	= C^\infty(\frF_\crit/\sfR)\,,
\end{equation}
where \(\sfR\) is the subgroup \(\Symm \geq \sfR \morneq \Gauge\) such that \(\sfR/\Gauge = \Gamma(\frR\times\frF/\cG)\). We denote their Lie algebra counterparts as \(\frr = \Lie(\frR) \leq \frs\) and \(\symm \geq \sfr = \Lie(\sfR) \morneq \gauge\) such that \(\sfr/\gauge = \Gamma(\frr\times\frF/\cG)\).

In particular, we wish to consider the case of \(\frR\) being a \emph{compact supertranslation group} of dimension \(\dim(\frR) = \sfb|1\), generated by a single supersymmetry \(\cQ \in \frr\), where
\begin{equation}
	\sfb
	=
	\left\{
	\begin{aligned}
		&0 & [\cQ,\cQ] &= 0
		&
		&\textsf{`nilpotent'}
		\\
		&1 & [\cQ,\cQ] &\eqcolon \cB \neq 0  
		&
		&\textsf{`equivariant'}
	\end{aligned}
	\right.
\end{equation}
where if non-vanishing, \(\cB\) generates a \(\operatorname U(1) = \frR_\circ\) action on \(\frF/\cG\). Note that for this supergroup \(\frR_\circ\normeq\frR\) and that \(\frR/\frR_\circ\) always corresponds to the \(\sfb = 0\) case.

Let us now denote such supersymmetric (i.e.~BPS) observables by \(\cO_\BPS \in C^\infty(\Omega)^\frR\). Their expectation values in the BRST formalism are then given by
\begin{equation}
	\EV{\cO_\BPS}
	\label{EV BRST}
	= \int_{\frF_\BRST} \extD\mu_\BRST ~ \cO_\BPS\mathrm e^{-S_\BRST}\,,
\end{equation}
where we do not normalise. The BRST action reads \(S_\BRST = S + Q_\BRST\Psi_\gf\), where \(\Psi_\gf\) is a gauge-fixing fermion and \(Q_\BRST\) the BRST differential on \(\frF_\BRST\) induced by the BV differential.

\subsection{Off-shell localisation argument rephrased using a localising fermion}
\label{second localisation argument}

We now review the usual argument for computing the expectation values \(\EV{\cO_\BPS}\) of BPS operators \(\cO_\BPS\) in the case where we have an off-shell (super)symmetry algebra, but rephrased in terms of the language of localising fermions to highlight the parallels to gauge fixing. For this approach we work in the BRST picture, i.e.\ expression \eqref{EV BRST}.

\paragraph{Lifting global symmetries to BRST space.} 

To introduce the localisation argument, we need to be able to describe global symmetries acting on \(\frF\) and \(\frF_\BRST\) rather than \(\frF/\cG\). Lifting the  global symmetries to \(\frF\) amounts to (non-canonically) choosing a section \(\Sigma \in \Gamma(\Symm\onto\Global)\) for finite gauge transformations, resp.\ a section \(\sigma \in \Gamma(\symm\onto\globall)\) at the infinitesimal level. Note, however, that these sections  are not generically group homomorphisms since the global symmetries may only close up to gauge symmetries.

Now, to lift the action of global symmetries from \(\frF\) to \(\frF_\BRST\): since we consider gauge symmetries forming  a normal subgroup \(\Gauge\normeq\Symm\), \(\Symm\) acts naturally on \(\Gauge\) via the adjoint action. Similarly, \(\symm\) acts naturally on \(\gauge\normeq\symm\) via the adjoint representation. It then follows that \(\symm\) acts in the coadjoint representation on ghost fields, i.e.\ a symmetry \(X\in\Gamma(\tT\frF)\) is lifted to \(\Gamma(\tT\frF_\BRST)\) through \(Xc^\alpha \coloneq -\ad_X^* c^\alpha\). As a consequence, the symmetries then commute with the BRST differential:
\begin{equation}
	[X,Q_\BRST] = 0\,,
\end{equation}
for every \(X\in\symm\).
In particular, representatives of global symmetries commute with the BRST differential.
Note that this lift from \(\Gamma(\tT\frF)\) to \(\Gamma(\tT\frF_\BRST)\) is \emph{not} necessarily a morphism of \(C^\infty(\frF)\)-modules.

After lifting symmetries to BRST space, they may not be an exact symmetry of the action (if the gauge-fixing procedure does not respect the underlying symmetries), but they remain a symmetry in the cohomology of the BRST differential:
\begin{equation}
\label{symm commutes with BRST}
	XS_\BRST
	=
	Q_\BRST\mleft((-)^{\|X\|}X\Psi_\gf\mright)\,,
\end{equation}
that is, \(X\) only annihilates \(S_\BRST\) up to a \(Q_\BRST\)-exact term.

\paragraph{Equivariant cohomology.}
The localisation argument uses techniques from equivariant cohomology. We start off by noting that if we take the global symmetries to be non-anomalous, \(\cQ\)-exact expressions do not contribute to the path integral. Indeed, in the absence of  global anomalies
\begin{equation}
	\div_{\extD\mu_{\frF/\cG}}\cQ = \div_{\extD\mu_{\frF/\cG}}\cB = 0\,,
\end{equation}
or equivalently, by lifting the symmetries from \(\frF/\cG\) to \(\frF_\BRST\),
\begin{align}
	\div_{\extD\mu_\BRST}\cQ = \div_{\extD\mu_\BRST}\cB &= 0
	&
	&\Leftrightarrow
	&
	\int\extD\mu_\BRST ~ \cQ(\cdots)
	=
	\int\extD\mu_\BRST ~ \cB(\cdots)
	&= 0\,.
\end{align}
One then finds that for a \(\cQ\)-exact observable \(\cQ\Xi\), where \(\Xi \in \ker Q_\BRST \subset C^\infty(\frF_\BRST)\), that its expectation value vanishes,
\begin{equation}
\label{Q comohology vanishing EV}
	\EV{\cQ\Xi}
	\overeq{\ref{symm commutes with BRST}}
	\int\extD\mu_\BRST ~ \Big[\cQ\big(\Xi\mathrm e^{-S_\BRST}\big) - Q_\BRST\big(\cQ\Psi_\gf \Xi\mathrm e^{-S_\BRST}\big)\Big]
	= 0\,.
\end{equation}
Restricted to the space of  \(\cB\)-equivariant observables \(H^\bullet(Q_\BRST)^{\frR_\circ} = \ker(\cB|H^\bullet(Q_\BRST))\), \(\cQ^2=0\). We then define \(\cB\)-equivariant \(\cQ\)-cohomology to be \(H^\bullet(\cQ|H^\bullet(Q_\BRST)^{\frR_\circ})\). In particular, this space defines an algebra, and this algebra is isomorphic to the space of `off-shell' BPS observables:
\begin{subequations}
\begin{align}
	H^0(Q_\BRST)^{\frR_\circ} 
	&= C^\infty(\frF/\cG)^{\frR_\circ}\,,
	\\
	\ker\big(\cQ\,\big|\,H^0(Q_\BRST)^{\frR_\circ}\big) 
	&= C^\infty(\frF/\cG)^\frR\,.
\end{align}
\end{subequations}
If we replace the kernel in the second line by the cohomology, we obtain the non-trivial BPS observables, that is, those which are not \(\cQ\)-exact.

\subsubsection{Localisation argument} 
\label{sssec:Localisation argument}

Let us now introduce a parity-odd gauge-invariant equivariant functional \(\Psi_\loc \in C^\infty_\odd(\frF)\),
\begin{align}
\label{localising fermion gauge invariance equivariance}
	Q_\BRST\Psi_\loc &= 0\,,
	&
	\cB\Psi_\loc &= 0\,,
\end{align}
which one should regard as the localisation analogue of the gauge-fixing fermion $\Psi_\gf$.

We observe that we are free to deform the action by a \(\cQ\)-exact term \(\cQ\Psi_\loc\) without changing the expectation value \(\EV{\cO_\BPS}\) of BPS operators \(\cO_\BPS\). Indeed, define
\begin{equation}
	\EV{\cO_\BPS}({t})
	\coloneq 
	\EV{\cO_\BPS\mathrm e^{{-t^2\cQ\Psi_\loc}}}
	=
	\int\extD\mu_\BRST ~ \cO_\BPS\mathrm e^{-S_\BRST {- t^2\cQ\Psi_\loc}}\,.
\end{equation}
Differentiating with respect to \(t \geq 0\) find then that 
\begin{equation}
	\tpder{}{t}\EV{\cO_\BPS}(t) 
	=
	\EV{-2t\cQ\big(\Psi_\loc \cO_\BPS\mathrm e^{-t\cQ\Psi_\loc}\big)}
	\overeq{\ref{Q comohology vanishing EV}}
	0\,,
\end{equation}
which is to say that we are free to deform the action as \(S_\BRST \to S_\BRST + t\cQ\Psi_\loc\), without changing expectation values of observables. If we can furthermore choose \(\Psi_\loc\) such that the deformation term has a positive semi-definite body,
\begin{equation}
	(\cQ\Psi_\loc)_\circ \geq 0\,,
\end{equation}
then we can sensibly take the limit \(t\to\infty\). In this limit, the path integral localises to the zero locus of the deformation term, which forms a subset of the localisation locus \(\frF_\loc\). We will henceforth refer to \(\Psi_\loc\) as the \emph{localising fermion}.

\paragraph{Constructing a localising fermion.} Now, to explicitly perform the localisation, let us take a closer look at the geometry of history space \(\frF\). We assume there to be a fermion number operator \(F \in \Gamma(\tT\frF)\) which lifts the \(\bbZ_2\)-grading on the supermanifold \(\frF\) to an \(\bbN\)-grading, for which the fermionic fields are assigned fermion number \(1\) (similar lifts appear in the twisting of field theories \cite{Costello:2011np,Elliott:2020ecf,Saberi:2021weg}).\footnote{In this context, the fermion number does not talk to the ghost number.} The space of histories \(\frF\) then obtains a canonical structure of a fermionic vector bundle \(\frF\to\frF_\circ\) over its body. Suppose that \(\tT\frF\) admits a splitting \(\tT\frF = \tT_0\frF \oplus \tT_1\frF\) into the vector bundles along bosonic and fermionic directions.
Henceforth, we will denote the collective bosons by \(\phi\) and the collective fermions by \(\lambda\), i.e.\ \(F = \lambda\delta/\delta\lambda\).
We take the supersymmetry to decompose into
\begin{align}
	\cQ
	&=
	\cQ_+ + \cQ_-\,,
	&
	\begin{aligned}
		\cQ_+ &\in \Gamma_{+1}(\tT_0\frF)
		\\
		\cQ_- &\in \Gamma_{-1}(\tT_1\frF)
	\end{aligned}
\end{align}
The analogy with equivariant localisation is \(\frF = \tT[1]M = (\Pi\tT M,F)\), where \(F\) is the form degree counting vector field, and \(\cQ = \extd_V\) is the equivariant differential decomposed into the de Rham differential \(\cQ_+ = \extd\) and interior product \(\cQ_- = \iota_V\).

Regarding \(\frF_\bbC\to\frF_\circ\) as a (complexified) vector bundle, we then assume that there exists an (antilinear) vector bundle involution \(\dagger : \frF_\bbC \to (\frF_\bbC)^\ast\), which defines a Hermitian vector bundle product which is invariant under gauge symmetries \(\cG\) and invariant under bosonic global symmetries \(\frR_\circ\). The degree \(-1\) component \(\cQ_-\) of the supersymmetry \(\cQ\) can then be interpret as a section \(\cQ_- \mapsto \cQ\lambda \in \Gamma(\frF,\bbC)\).\footnote{Henceforth, we will assume that we work over complexified spaces and rings of functions, without necessarily explicitly writing this down.} We can then write down a localising fermion 
\begin{equation}
	\Psi_\loc(t)
	=
	\tfrac12 t^2(\cQ\lambda)^\dagger\lambda
	\in 
	C^\infty(\frF,\bbC)\,.
\end{equation}
This localising fermion then satisfies the properties
\begin{align}
	(\cQ\Psi_\loc)_\circ
	&= 
	\tfrac12 t^2(\cQ\lambda)^\dagger(\cQ\lambda) 
	\geq 0\,,
	&
	Q_\BRST\Psi_\loc &= 0\,,
	&
	\cB\Psi_\loc &= 0\,.
\end{align}
Thus, it is suitable for localisation. In the limit \(t\to\infty\) the path integral now localises to the zeroes \(\iota_\loc\colon \frF_\loc \coloneq \ker(\cQ_-) \into \frF_\circ\) of \(\cQ_-\). The contributions to the path integral then localise to the infinitesimal region around the localisation locus \(\frF_\loc\), given by the normal bundle \(\tN\frF_\loc \coloneq \iota_\loc^\ast\tT\frF/\tT\frF_\loc\) (i.e.\ at the locus \(\frF_\loc\), we take the quotient of \emph{all} tangent vectors of \(\frF\) by those that are tangent to \(\frF_\loc\)). What precisely we take \(\frF\) to be for a gauge theory will be specified in subsequent paragraphs.
We then integrate out the fibres and are left with an integral over \(\frF_\loc\). We note that in supersymmetric theories, the normal vectors organise themselves into boson--fermion doublets, which can be canonically integrated out. This is to say, the measure on \(\frF\) induces a canonical measure on \(\tN\frF_\loc\), and canonically\footnote{This is because the parity-reversed tangent bundle \(\Pi\mathrm{T}\mathcal{M}\) of a supermanifold \(\mathcal M\) has a canonical measure \([\mathrm dx|\mathrm d\xi]\), where \(x^\mu\) are the coordinates and \(\xi^\mu\) are the fibre coordinates. In this context, taking the normal fibres to organise into doublets is tantamount to saying the typical fibre is a linear supermanifold of this form.} integrating out the doublets along the fibre leaves us with a measure on \(\frF_\loc\).

\paragraph{Mixed approach to localisation.} Before we localise we first impose gauge-fixing conditions. Suppose we wish to impose gauge-fixing conditions \(\chi_\alpha[\phi] \in C^\infty(\frF)\), where \(\alpha = \alpha_0\) is the \(\frg = \frg_0\) DeWitt index. We take these to fully fix the gauge in the sense that\footnote{We use `rank' and `dimension' somewhat loosely in this setting since we work with infinite-dimensional spaces. Roughly speaking, we take this to mean that the local degrees of freedom match.}
\begin{align}
\label{conditions on gauge fixing conditions}
	\rank \big(\fder{}{\phi^i}\chi_\alpha\big) &= \dim \frg\,,
	&
	\rank \big(R_\alpha{^i}\fder{}{\phi^i}\chi_\beta\big) &= \dim \frg\,,
\end{align}
where \(R_\alpha \in \Gamma(\tT\frF)\) are the gauge transformations. One then introduces a trivial pair \((\bar{c}^\alpha,b^\alpha)\) consisting of an antighost \(\bar{c}^\alpha\) and a Nakanishi--Lautrup field \(b^\alpha\) of bidegrees 
\begin{align}
	|\bar{c}| &= (-1,\even)\,,&
	|b| &= (0,\even)\,.
\end{align}
The gauge-fixing fermion is now given by
\begin{align}
\label{eq:R xi gauge fixing fermion}
	\Psi_\gf
	&=
	\bar{c}^\alpha\big(\tfrac{\xi}{2}\kappa_{\alpha\beta}b^\beta + \chi_\alpha\big)
	&
	&\left\{
	\begin{aligned}
		Q_\BRST\bar{c}^\alpha &= \rmi b^\alpha
		\\
		Q_\BRST b^\alpha &= 0
	\end{aligned}
	\right.\,,
\end{align}
where $\xi \geq 0$ and \(\kappa_{\alpha\beta}\) is a non-degenerate gauge-invariant bilinear form. For simplicity, we consider the case \(\xi = 0\), corresponding to a \(\delta\)-function gauge, as opposed to an \(R_\xi\)-gauge. The BRST action is then given by
\begin{align}
	S_\BRST
	&=
	S + S_\gf\,,
	&
	S_\gf
	&\coloneq 
	Q_\BRST\Psi_\gf
	=
	\rmi b^\alpha\chi_\alpha
	+
	c^\alpha(R\chi)_{\alpha\beta}\bar{c}^\beta\,,
\end{align}
and the path integral measure takes on the form
\begin{equation}
	\extD\mu_\BRST
	=
	\extD\phi\,\extD\lambda\,\extD c\,\extD\bar c\,\extD b
	\xrightarrow{\int\extD b}
	(\extD\phi_\gf\,\extD\lambda)(\extD c\,\extD\bar c)
	\eqcolon
	\extD\mu_\gf\,\extD\mu_{\gh}\,,
\end{equation}
where we integrated out the Nakanishi--Lautrup field, which acts as a Lagrange multiplier to enforce the gauge conditions. Restricting ourselves to the localisation locus, the path integral measure \(\extD\mu_\gf\) now factorises into
\begin{equation}
	\iota_\loc^\ast\extD\mu_\gf
	=
	(\extD\phi_\loc^\gf\,\extD\lambda)(\extD\delta\phi_\gf)
	\eqcolon
	\extD\mu_\loc^\gf\,\extD\mu_\perp\,.
\end{equation}
Here, \(\phi^\gf_\loc\) are the coordinate functionals of the gauge fixed localisation locus \(\frF_\loc^\gf\) and \((\delta\phi_\gf,\lambda)\) are the linear coordinate functionals of the normal fibres, forming supersymmetry doublets. Finally, we perform a rescaling
\begin{align}
\label{eq:theory rescaling mixed approach}
	&
	\begin{aligned}
		\delta\phi_\gf &= \tfrac{1}{t}\bar\delta\phi_\gf
		\\
		\lambda &= \tfrac{1}{t}\bar\delta\lambda
	\end{aligned}\,,
	&
	\extD\lambda\,\extD\delta\phi_\gf
	&= 
	\Ber\fder{(\phantom{\bar\delta}\lambda,\delta\phi_\gf)}{(\bar\delta\lambda,\bar\delta\phi_\gf)}
	\extD\bar\delta\lambda\,\extD\bar\delta\phi_\gf
	=
	\extD\bar\delta\lambda\,\extD\bar\delta\phi_\gf\,.
\end{align}
The Jacobian does not contribute to the transformation of the measure: gauge-fixed bosonic and fermionic modes around the locus lie in supersymmetry doublets, so that the rescalings contribute with opposing factors to the Berezinian, cancelling each other out.

Everything is now in place for localisation. Performing the path integral, we find that
\begin{equation}
\begin{split}
	\EV{\cO_\BPS}
	&= \lim_{t\to\infty}\EV{\cO_\BPS}(t)
	\\
	&= \lim_{t\to\infty}\int_{\frF_\BRST} \extD\phi\,\extD\lambda\,\extD c\,\extD\bar c\,\extD b ~ \cO_\BPS\rme^{-S - S_\gf - t^2S_\loc}
	\\
	&= \lim_{t\to\infty}\int_{\frF^\gf_\BRST}\extD\phi_\gf\,\extD\lambda\,\extD c\,\extD\bar{c} ~ \cO_\BPS\rme^{-S - S_\gf - t^2S_\loc}
	\\
    &= \lim_{t\to\infty}\int_{\tN\frF_\loc^\gf}\extD\phi_\loc^\gf( \extD\lambda\,\extD\delta\phi_\gf)(\extD c\,\extD\bar c) ~ \cO_\BPS\rme^{-S - S_\gf - t^2S_\loc}
	\\
	&= \lim_{t\to\infty}\int_{\tN\frF_\loc^\gf}\extD\phi_\loc^\gf( \extD\bar\delta\lambda\,\extD\bar\delta\phi_\gf)(\extD c\,\extD\bar c) ~ \cO_\BPS \rme^{-S - S_\gf^\lin - S_\loc^\lin} + \Order(t^{-1})
	\\
	&= \int_{\frF_\loc^\gf}\extD\phi^\gf_\loc \frac{\cO_\BPS\rme^{-S}}{\Ber^{\frac12}\Hess_\perp(S_\loc + S_\gf)}
	\,,
\end{split}
\end{equation}
where \(\lin\) is used to denote the linearised action, and \(\Hess_\perp\) denotes the Hessian along the directions normal to the locus.
This is what we may coin the \emph{mixed approach} to localisation, and is typically what is done in the literature (cf.\ \cite{Kapustin:2009kz}). The reason for this wording is that fields, ghosts and trivial pairs are not treated on equal footing in this approach to localisation. Now, let us present what we coin the \emph{democratic approach} to localisation, in which we treat fields, ghosts and trivial pairs on the same footing.

\paragraph{Democratic approach to localisation.} In the democratic approach, we do \emph{not} start by integrating out the Nakanishi--Lautrup field. Instead, we regard the whole of BRST configuration space as the field space we are interested in. In this spirit, we regard the gauge-fixing term as a contributor to the localisation, and deform the BRST action as
\begin{align}
	&
	\begin{aligned}
		S_\BRST 
		&\to S_\BRST(t) + S_\loc(t) 
		\\
		&= S + S_\gf(t) + S_\loc(t)
	\end{aligned}\,,
	&
	&\left\{
	\begin{aligned}
		S_\gf(t)
		&= (1+t^2)Q_\BRST\Psi_\gf
		\\
		S_\loc(t) 
		&= t^2\cQ\Psi_\loc
	\end{aligned}\,,
	\right.
\end{align}
In the limit \(t\to\infty\) the path integral now instead localises to \emph{the gauge fixing conditions and the localisation locus.} The BRST configuration measure then decomposes on the gauge-fixed localisation locus as
\begin{equation}
	(\iota_\loc^\gf)^\ast\extD\mu_\BRST
	=
	(\extD\phi_\loc^\gf\,\extD\lambda\,\extD c\,\extD\bar{c})(\extD\delta\phi\,\extD b)
	\eqcolon \extD\mu_\loc^\gf\,\extD\mu_\perp\,,
\end{equation}
where we point out that now the fluctuations \(\delta\phi\) around the gauge-fixed locus \(\frF_\loc^\gf\) are not gauge-fixed themselves because we have not integrated out the Nakanishi--Lautrup field \(b\). Similarly, we can again rescale the fluctuations around the localisation locus as
\begin{align}
	&
	\begin{aligned}
		\delta\phi &= \tfrac{1}{t}\bar\delta\phi\,,
		&
		\lambda &= \tfrac{1}{t}\bar\delta \lambda\,,
		\\
		&
		&
		c &= \tfrac{1}{t}\bar\delta c\,,
		\\
		b &= \tfrac{1}{t}\bar\delta b\,,
		&
		\bar c &= \tfrac{1}{t}\bar\delta\bar c\,,
	\end{aligned}
	&
	\extD\lambda\,\extD c\,\extD\bar{c}\,\extD\delta\phi\,\extD b
	&=
	\extD\bar\delta\lambda\,\extD\bar\delta c\,\extD\bar\delta\bar{c}\,\extD\bar\delta\phi\,\extD\bar\delta b\,,
\end{align}
Again, the Jacobian factor cancels out after this rescaling because the non-gauge-fixed fluctuations \(\delta\phi\) --- which split up in gauge-fixed and pure gauge modes --- form boson--fermion doublets with the fermions \(\lambda\) and the ghosts \(c\), and the trivial pair \((\bar c,b)\) also lies in an boson--fermion doublet.

Performing the path integral, we now find
\begin{equation}
\begin{split}
	\EV{\cO_\BPS}
	&= \lim_{t\to\infty}\EV{\cO_\BPS}(t)
	\\
	&= \lim_{t\to\infty}\int_{\frF_\BRST} \extD\phi\,\extD\lambda\,\extD c\,\extD\bar c\,\extD b ~ \cO_\BPS \rme^{-S - (1 + t^2) S_\gf - t^2S_\loc}
	\\
	&= \lim_{t\to\infty}\int_{\tN\frF_\loc^\gf}
	\extD\phi_\loc^\gf(\extD\lambda\,\extD c\,\extD\bar c\,\extD\delta\phi\,\extD b) ~ \cO_\BPS \rme^{-S - (1 + t^2) S_\gf - t^2S_\loc}
	\\
	&= \lim_{t\to\infty}\int_{\tN\frF_\loc^\gf}
	\extD\phi_\loc^\gf(\extD\bar\delta\lambda\,\extD\bar\delta c\,\extD\bar\delta\bar c\,\extD\bar\delta\phi\,\extD\bar\delta b) ~ \cO_\BPS \rme^{-S - S_\gf^\lin - S_\loc^\lin} + \Order(t^{-1})
	\\
	&= \int_{\frF^\gf_\loc}\extD\phi^\gf_\loc ~ \frac{\cO_\BPS e^{-S}}{\Ber^{\frac12}\Hess_\perp(S_\loc + S_\gf)}
\end{split}
\end{equation}
We obtain the same result again, but now, we gain a new perspective on the way the localising term and gauge-fixing term combine. One perspective on this is that localisation and gauge-fixing are actually very similar in nature. And indeed, one way of looking at localisation is as gauge-fixing with respect to a global symmetry. That is, we reduce the path integral to an integral over orbits of the symmetries in consideration. When this symmetry is a supersymmetry, only the orbits of singlet configurations will contribute because the doublets cancel each other out, much like how ghosts cancel out pure gauge modes.

\subsection{BRST reformulation of the off-shell localisation argument}
\label{ssec:BRST reformulation of the off-shell localisation argument}

\subsubsection{BRST formulation of global symmetries}
The aforementioned result can be recast in the language of the BRST/BV formalism, by treating global and gauge symmetries on the same footing in an extended BRST formalism. Consider the degree-shifted Lie algebroid
\begin{equation}
	\frF_\gBRST
	\coloneq 
	\frr[1] \times \frF_\BRST
\end{equation}
where one takes \(\frr \leq \frs\) to be a subalgebra of the global symmetry algebra, which we take to be the compact \(\sfb|1\)-dimensional supertranslation algebra. The homological vector field on this Lie algebroid is given by the global--gauge BRST operator
\begin{align}
	Q_\gBRST
	&=
	Q_\BRST
	+
	Q + B + Q_\CE^\frr\,,
	&
	&\left\{
	\begin{aligned}
		Q &= \varepsilon\cQ
		\\
		B &= \xi\cB
		\\
		Q_\CE^\frr
		&= -\varepsilon^2\tpder{}{\xi}		
	\end{aligned}\,,
	\right.
\end{align}
where \(\varepsilon, \xi \in C_\bullet^\infty(\frF_\gBRST)\) are the respectively even and odd linear coordinate functions on \(\frr[1]\), dual to  \(\cQ\) and \(\cB\), respectively, which we will refer to as \emph{global ghosts}. They carry \((\bbZ\times\bbZ_2)\)-valued bidegrees
\begin{align}
    |\varepsilon| &= (+1,\odd)\,,&
    |\xi| &= (+1,\even)\,.
\end{align}
These play a role similar to the formal variable \(u\) used to compensate degrees during twisting in \cite[§2.1]{Saberi:2021weg}. The key conceptual point here is that $\cQ$ carries cohomological degree $0$, so that a twist is required for it be regarded as a BRST operator. 

In the present BV--BRST context, the twisting procedure directly corresponds to the familiar BRST operation of replacing a gauge transformation $\theta$ into a ghost field $c$. Here, if the supersymmetry transformation parameter is denoted $\epsilon$, we have
\begin{equation}
\delta_\epsilon \Phi = \epsilon \cQ \Phi \quad \longrightarrow \quad Q \Phi = \varepsilon \cQ \Phi,
\end{equation}
the only conceptual difference being that, since we are applying this to a global symmetry, the ghost field $\varepsilon$ will be constant.

The extended BRST differential describes the `off-shell' BPS observables in its degree-\(0\) cohomology:
\begin{equation}
	H^0(Q_\gBRST)
	=
	C^\infty(\frF/\cG/\frR)\,.
\end{equation}

For purposes to be explained in the following, we will `turn on' the supersymmetries by restricting  to the subsupermanifold
\begin{equation}
	\frF_\gBRST^{\+} \coloneq \frr[1]^{\+} \times \frF_\BRST\,.
\end{equation}
Here, we define \(\frr[1]^{\+}\) to be the \(\bbZ\)-graded supermanifold given by
\begin{align}
	\frr[1]^{\+}
	&\coloneq
	\big(\Pi\frr^{\+},\Upsilon|_{\Pi\frr^{\+}}\big)\,,
	&
	\Pi\frr^{\+} = \bbC^\times \times \Pi\bbC^\sfb \subset \Pi\frr\,,
\end{align}
where \(\Pi\frr\) is the underlying supermanifold of \(\frr[1]\) and \(\Upsilon\) is the Euler vector field of \(\frr[1]\). 
This effectively amounts to enlarging the ring of functions in consideration, by giving the supersymmetry global ghost a \emph{multiplicative inverse} \(\tfrac1\varepsilon \in C^\infty(\Pi\frr^{\+})\).
That is, the graded rings of functions on \(\frF_\gBRST\) and \(\frF_\gBRST^{\+}\) are given by polynomials
\begin{align}
	C_\bullet^\infty(\frF_\gBRST) &= C_\bullet^\infty(\frF_\BRST)[\varepsilon,\xi]\,,
	&
	C_\bullet^\infty(\frF_\gBRST^{\+}) &= C_\bullet^\infty(\frF_\BRST)[\varepsilon,\tfrac{1}{\varepsilon},\xi]\,.
\end{align}
Accordingly, there is a natural inclusion \(C^\infty_\bullet(\frF_\gBRST)\into C_\bullet^\infty(\frF_\gBRST^{\+})\).\footnote{Here, \(C^\infty_{(k,p)}(\frF_\gBRST^\times) = \bigoplus_{\ell = 0}^\infty\varepsilon^{-\ell} C_{(k+\ell,p+[\ell])}^\infty(\frF_\gBRST)\), where we do \emph{not} allow for formal power series in this decomposition.}
We give explicit examples of the antighost and Nakanishi--Lautrup field construction in sections \ref{ssec:R xi gauge localisation superparticle} and \ref{ssec:R xi gauge localisation SYM}, and will develop the general theory in  subsequent work.

\subsubsection{Global--gauge BRST action via combined gauge-fixing and localising fermion}\label{sssec:loc-gf-BRST}

Consider now the global--gauge BRST action associated to the extended symmetry structure, 
\begin{align}
	S_\gBRST 
	&= 
	S + Q_\gBRST\Psi(t)\,,
	&
	\Psi(t) &\in C^\infty_{(-1,\even)}(\frF_\gBRST^\+)\,,
\end{align}
with \(t\geq 0\) and where we take the \emph{combined fermion} \(\Psi(t)\) to combine the gauge-fixing fermion \(\Psi_\gf\) and localising fermion \(\Psi_\loc\) into
\begin{equation}\label{gauge fixing localising fermion}
	\Psi(t)
	\coloneqq
	\Psi_\gf + \frac{t^2}{\varepsilon}\Psi_\loc\,.
\end{equation}
We assume the following invariances:
\begin{align}
	\cB\Psi_\loc &= 0\,,
	&
	Q_\BRST\Psi_\loc &= 0\,,
	&
	Q_\CE^\frr\Psi_\loc &= 0\,,
	&
	Q_\CE^\frr\Psi_\gf &= 0\,.
\end{align}
The first two conditions are the gauge invariance and equivariance conditions  \eqref{localising fermion gauge invariance equivariance} of the localising fermion, whereas the latter two conditions simply state that neither the localising fermion nor the gauge-fixing fermion may depend on \(\xi\). Assuming these invariances, the global--gauge BRST action becomes
\begin{equation}\label{eq:BRSTaction}
	S_\gBRST
	=
	S + \underbrace{Q_\BRST\Psi_\gf}_{\mathclap{\text{gauge-fixing term}}} 
	+ \underbrace{(\varepsilon\cQ + \xi\cB)\Psi_\gf}_{\text{mixing terms}} 
	+ \underbrace{t\cQ\Psi_{\loc\vphantom{\textsf{g}}}}_{\mathclap{\text{localising term}}}.
\end{equation}
The first term is the gauge-fixing term; the last term is the localising term\footnote{In deriving \eqref{eq:BRSTaction}, we assume that the globally extend BV action, $S_\gBV$ does not depend on the antifield global ghost $\varepsilon^+$. Otherwise, it may be the case that, for instance, $Q_\gBRST \frac{t^2}{\varepsilon}\Psi_\loc = -\frac{t}{\varepsilon^2}\Psi_\loc+\cdots$.  This assumption always holds for the examples of global symmetries considered, since $[\cQ,\cQ]\sim \cB$, so that the BV action may have  $\xi^+$ dependence, but no $\varepsilon^+$ dependence.}. The second term, however,  mixes global symmetries with the gauge-fixing fermion. This term compensates for the fact that the gauge-fixing condition may not be invariant under the global symmetries.

The mixing terms can be absorbed into a field redefinition, provided the gauge is fully fixed, as follows.
Let \(X = X^i\tfder{}{\phi^i} \in\Gamma(\tT_\frF\frF_\gBRST^\+)\) now be an arbitrary vector field along the \(\frF\)-directions. A consequence of the conditions \eqref{conditions on gauge fixing conditions} is that changes along this vector field along the gauge conditions can be absorbed into gauge transformations in the sense that there exists a vector field
\begin{equation}
	\Theta \coloneqq \Theta^\alpha\tfder{}{c^\alpha} \in \Gamma(\tT_{\frg[1]}\frF_\gBRST^\+)
\end{equation}
such that
\begin{equation}
	X^i\fder{\chi_\alpha}{\phi^i} = 
	\Theta^\beta(R\chi)_{\beta\alpha}\,.
\end{equation}
Suppose we now shift the ghost fields \(c^\alpha\) by the section \(\Theta^\alpha\) for \(X = \varepsilon\cQ + \xi\cB\). Then the mixing terms are absorbed by this field redefinition into the gauge-fixing term:
\begin{align}
	c^\alpha 
	&\mapsto 
	c^\alpha - \Theta^\alpha
	&
	&\Longrightarrow
	&
	S_\gBRST &\mapsto S + Q_\BRST\Psi_\gf + t\cQ\Psi_\loc\,.
\end{align}

\subsubsection{Localisation of the path integral}
Given the gauge-fixing/localising fermion, we can now  address the path integral.
Since we do not gauge the global symmetries, the path integral on the global--gauge BRST space becomes an operation
\begin{equation}
	\int_{\frF_\BRST}\extD\mu_\BRST \colon C^\infty(\frF_\gBRST^\+) \to C_\bullet^\infty(\frr[1]^\+) \eqqcolon \CE(\frr)^\+\,,
\end{equation} 
where \(\CE(\frr)^\+\) is the (complexified) Chevalley--Eilenberg algebra \(\CE(\frr)\) ring-theoretically localised to \(\varepsilon \neq 0\). To ensure that the result is physically sensible, it needs to lie in the subring \(\bbC \subset C^\infty(\frr[1]^\+)\). The space of such functions is given by the degree-\(0\) kernel of the Chevalley-Eilenberg differential \(Q_\CE^\frr = -\varepsilon^2\tpder{}{\xi}\),
\begin{equation}
	\bbC = \ker\big(Q_\CE^\frr\,\big|\,\CE_0(\frr)^\+\big)\,.
\end{equation}
Note that this differs from the cohomology since \(H^0(Q_\CE^\frr|\CE(\frr)^\+) = 0\). Consider now a BPS operator \(\cO_\BPS \in C^\infty(\frF_\BRST)\). We can relate its value given in the BRST formalism to the global--gauge BRST formalism by reverse shifting the ghost \(c^\alpha \to c^\alpha + \Theta^\alpha\) to give
\begin{align}
	&
	\begin{aligned}
		\EV{\cO_\BPS}
		&= \int_{\frF_\BRST}\extD\mu_\BRST ~ \cO_\BPS\mathrm e^{-S - Q_\BRST\Psi_\gf}
		\\
		&= \int_{\frF_\BRST}\extD\mu_\BRST ~ \cO_\BPS\mathrm e^{-S - Q_{{\gBRST}}\Psi_\gf}\,.
	\end{aligned}
\end{align}
Note that the would-be  Jacobian factor following from \(c^\alpha \to c^\alpha + \Theta^\alpha\)  is  unity,
\begin{equation}
\Ber\fder{}{c^\beta}(c^\alpha + \Theta^\alpha)
	=
	\Ber\delta^\alpha{_\beta}
	=
	1.
\end{equation}

We introduce the localising fermion through \(t\)-dependent term as
\begin{align}
	\EV{\cO_\BPS}(t) 
	&=
	\int_{\frF_\BRST}\extD\mu_\BRST ~ \cO_\BPS\mathrm e^{-S - Q_\gBRST\Psi(t)}\,,
	&
	\Psi(t) &= \Psi_\gf + \frac{t^2}{\varepsilon}\Psi_\loc\,.
\end{align}
Crucially, the path  integral is \(t\)-independent for an appropriate choice of localising fermion. Taking the \(t\)-derivative we find that
\begin{equation}
\label{BRST BPS equivariance conditions}
\begin{split}
	\pder{}{t}\EV{\cO_\BPS}(t)
	&= -\int_{\frF_\BRST}\extD\mu_\BRST ~ (Q_\gBRST\dot\Psi)\cO_\BPS\mathrm e^{-S-Q_\gBRST\Psi}
	\\
	&= 
	\begin{aligned}[t]
		&- \int_{\frF_\BRST}\extD\mu_\BRST ~ \dot\Psi \big(Q_\gBRST\cO_\BPS\big)\mathrm e^{-S-Q_\gBRST\Psi}
		&
		&\Big\}{\textsf{ BPS}}
		\\
		&- Q_\CE^\frr\int_{\frF_\BRST}\extD\mu_\BRST ~ \dot\Psi\cO_\BPS\mathrm e^{-S - Q_\gBRST\Psi}\,.
		&
		&\Big\}{\textsf{ equivariance}}
	\end{aligned}
\end{split}
\end{equation}
To arrive at the second step we use the Leibniz rule, together with the fact that \(Q_\BRST\), \(\cQ\) and \(\cB\) are taken to be non-anomalous with respect to the path integral measure \(\extD\mu_\BRST\). We require  \(\cO_\BPS\) to be \(t\)-independent. The first line in the second step imposes the BPS condition on \(\cO_\BPS\), and the second line, when unpacked, imposes equivariance conditions and gauge invariance on the localising fermion \(\Psi_\loc\).
Following this step, all of the computations that follow are the same.

\section{Localisation for on-shell algebras in the Batalin--Vilkovisky formalism}\label{sec:on shell localisation}

\subsection{Batalin--Vilkovisky treatment of global symmetries}
The first step towards a BV enhancement of the localisation argument is to treat global and gauge symmetries on an equal footing in the BV formalism. To this end, let us consider a subalgebra \(\sfr \leq \globall \leq \Gamma(\tT\Omega)\). Note that taking the preimage of the projection \(\pi\colon \symm \to \globall = \symm/\gauge\) corresponds to an enlargement \(\pi^{-1}(\sfr) \geq \gauge\) of the gauge group.

Following techniques from homological perturbation theory, one constructs the global--gauge BV space \(\frF_\gBV\) using precisely the same steps. Accordingly, it will be the degree-shifted cotangent bundle of global--gauge BRST space, \(\frF_\gBV = \tT^*[-1]\frF_\gBRST\). We already denoted the level-\(k\) gauge symmetry ghosts by \(c^{\alpha_k}\). Let us denote the level-\(k\) global symmetry ghosts by \(\Xi^A = (\xi^{a_0},\xi^{a_1},\xi^{a_2},\cdots)\). Using homological perturbation theory, one may then construct a global--gauge BV differential \(Q_\gBV \in \Gamma_{(+1,\even)}(\tT\frF_\gBV)\), which describes the \(\sfR\)-invariant observables in its degree-\(0\) cohomology\footnote{When we don't specify the fermion parity in the cohomology, we take this to mean that the cohomology can lie in either fermion parity.}
\begin{equation}
	H^0(Q_\gBV) 
	= 
	C^\infty\big(\frF_\crit\big/\pi^{-1}(\sfR)\big)
	=
	C^\infty(\Omega)^\sfR\,,
\end{equation}
where \(\sfR\) the Lie group obtained by integrating \(\sfr = \Lie(\sfR)\). We then extend the BV symplectic form \(\omega_\BV\), BV bracket \((-,-)_\BV\) and BV Laplacian \(\Delta_\BV\) introduced in \eqref{BV symplectic form etc} to their global--gauge BV analogues
\begin{subequations}
\begin{align}
	\omega_\gBV
	&= \omega_\BV + \omega_\frr\,,
	&
	\omega_\frr &= (-)^{\|\Xi^A\|}\extd\Xi^A\wedge\extd\Xia_A
	\bigg.\\
	(F,G)_\gBV 
	&= (F,G)_\BV + (F,G)_\frr\,,
	&
	(F,G)_\frr &= F\Big(\pderr{}{\Xia_A}\pderl{}{\Xi^A} - \pderr{}{\Xi^A}\pderl{}{\Xia_A}\Big)G
	\bigg.\\
	\Delta_\gBV
	&= \Delta_\BV + \Delta_\frr\,,
	&
	\Delta_\frr &= (-)^{\|\Xi^A\|}\pder{}{\Xi^A}\pder{}{\Xia_A}
	\bigg.
\end{align}
\end{subequations}
Here, we use finite-dimensional as opposed to functional notation for the global-ghost derivatives since typically the indices \(A = (a_0,a_1,a_2,\dotsc)\) will be taken to be discrete; however this is mostly just a matter of notation.
Consistent with \(\cL_{Q_\gBV}\omega_\gBV = 0\) the global--gauge BV differential \(Q_\gBV = (S_\gBV,-)\) is taken to be Hamiltonian with respect to the global--gauge BV action \(S_\gBV \in C^\infty_{(0,\even)}(\frF_\gBV)\). Nilpotence then implies the \emph{classical global--gauge master equation}
\begin{align}
\label{classical ghobal master equation}
	Q_\gBV^2 &= 0
	&
	&\Leftrightarrow
	&
	(S_\gBV,S_\gBV)_\gBV &= 0\,.
\end{align}
We require that the solution \(S_\gBV\) to \eqref{classical ghobal master equation} is proper, as defined in \cite[\S{}4.3]{Gomis:1994he} and \cite[\S{}17.3.2]{Henneaux:1994lbw}\footnote{Specifically, requiring that the solution is proper is to impose boundary conditions which guarantee that the cohomology of the (g)BV complex describes the correct space of observables. An example of a solution to the (global--gauge) master equation which is not proper is \(S_{(\mathsf g)\BV} = 0\).}. It is then unique up to canonical transformations and inclusion of trivial pairs.

\subsection{Quantisation of the BV theory with global ghosts}
To quantise the global--gauge BV theory, we take the global--gauge BV measure \(\extD\mu_\gBV\) to decompose into
\begin{subequations}
\begin{align}
	\extD\mu_\gBV &= \extd\mu_\frr\,\extD\mu_\BV\,,
	&
	\extd\mu_\frr &= \extd\Xi\,\extd\Xia\,,
	\\
	\extD\mu_\gBRST \coloneq \sqrt{\extD\mu_\gBV} &= \sqrt{\extd\mu_\frr}\,\sqrt{\extD\mu_\BV} = \sqrt{\extd\mu_\frr}\,\extD\mu_\BRST\,,
	&
	\sqrt{\extd\mu_\frr} &= \extd\Xi\,.
\end{align}
\end{subequations}
By restricting to a subspace \(\frr[1]^\+\into\frr[1]\), we may then consider subspaces \(\frF_\gBRST^\+\) and \(\frF_\gBV^\+\). The path integral then corresponds to an operation
\begin{equation}
	\int_{\cL_\Psi}\sqrt{\extD\mu_\BV}
	=
	\int_{\cL_\Psi}\sqrt{\extD\mu_\gBV}\,\delta_\Xi
	: 
	C^\infty(\frF_\gBV^\+) 
	\xlongrightarrow{\cL_\Psi} C^\infty(\frF_\gBRST^\+) 
	\xlongrightarrow{\int} C_\bullet^\infty(\frr[1]^\+)\,.
\end{equation}
The would-be integral over the global ghosts \(\Xi^A\) is circumvented by inserting a delta function \(\delta_\Xi\) on \(\frr[1]^\+\) concentrated in the global-ghost coordinate functions \(\Xi^A \in C^\infty(\frr[1]^\+)\). Consequently, the path integral will  generically  depend on the fermion \(\Psi\in C^\infty_{(-1,\even)}(\frF_\gBRST^\+)\) through its $\Xi$ dependence. Therefore, given \(\cO_\BPS \in C^\infty_0(\frF_\gBV^\+)\), let us define a functional
\begin{align}
	\EV{\cO_\BPS}[{\Psi}]
	&\coloneq 
	\int_{\cL_{{\Psi}}}\sqrt{\extD\mu_\BV} ~ \cO_\BPS\mathrm e^{-S_\gBV}
	&
	&:
	&
	C_{(-1,\even)}^\infty(\frF_\gBV^+)
	&\to 
	C^\infty_{(0,\even)}(\frr[1]^+)
\end{align}
with a smooth family of degree-\((-1)\) fermions \(\Psi(t)\). We then find that
\begin{equation}
\label{BV BPS equivariance conditions}
\begin{split}
	\pder{}{t}\EV{\cO_\BPS}[\Psi(t)]
	&=
	\int_{\cL_\Psi}\sqrt{\extD\mu_\gBV} ~ \tpder{}{t}\Psi\Delta_\gBV\Big\{\delta_\Xi\cO_\BPS\mathrm e^{-S_\gBV}\Big\}
	\\
	&= 
	\begin{aligned}[t]
		&\phantom{{}+{}}\int_{\cL_\Psi}\sqrt{\extD\mu_\BV} ~ \tpder{}{t}\Psi{\Delta_\gBV\Big\{\cO_\BPS\mathrm e^{-S_\gBV}\Big\}}
		&
		&\Big\}{\textsf{ BPS}}
		\\
		&- \int_{\cL_\Psi}\sqrt{\extD\mu_\BV} ~ {\Delta_\frr\Big\{\tpder{}{t}\Psi\cO_\BPS\mathrm e^{-S_\gBV}\Big\}}
		&
		&\Big\}{\textsf{ equivariance}}
	\end{aligned}
\end{split}
\end{equation}
These steps are the BV analogue of the BRST manipulations given in \eqref{BRST BPS equivariance conditions}. Accordingly, the first line in the second step expresses that \(\cO_\BPS\) be a quantum \(\sfR\)-invariant observable, while when one unpacks the second line one obtains the equivariance conditions.

We remark that, while this argument gives us a way of introducing a localising fermion in the on-shell setting, it does not yet give us a canonical choice of localising fermion or a guarantee that the deformation will localise; the literature treats this issue on an ad~hoc, case-by-case basis.

\subsection{Ansätze for solutions to the master equation}
We now briefly discuss which form the solutions to the master equation will take on for use in the later examples (sections~\ref{sec:superparticle} and \ref{sec:seifert}).

\paragraph{Closed Lie algebra representation without gauge symmetry.} For the off-shell supermultiplet, the global supersymmetry algebra forms an honest Lie (super)algebra representation. Let us denote the generators by \(T_a = T_a{^i}\fder{}{\phi^i}\in \Gamma(\tT\frF)\), and \(f_{ab}{^c}\) the structure constants. These then satisfy
\begin{align}
	[T_a,T_b] &= f_{ab}{^c}T_c\,,
	&
	f_{[ab}{^d}f_{c]d}{^e} &= 0\,.
\end{align}
In this setting, the minimal proper solution to the global--gauge master equation is given by
\begin{equation}
\label{off-shell BV solution}
\begin{split}
	S_\gBV
	&=
	S + \xi^a T_a{^i}\phia_i - \tfrac12 (-)^{(\xi^a+1)\xi^b}\xi^a \xi^bf_{ab}{^c}\xi^+_c
	\\
	&= S + \xi^a T_a^\asta - Q_\CE^\asta\,,
\end{split}
\end{equation}
where we write \((-)^F \coloneq (-)^{\|F\|}\) and for a vector field $V= V^I\frac{\delta}{\delta \Phi^I}$, $V^\asta \coloneq V^I\Phi^+_I$.

\paragraph{Open Lie algebra representation of the bivector type without gauge symmetry.} For the on-shell supermultiplet, the representation will cease to close off-shell. It then takes on the form
\begin{align}
	[T_a,T_b] 
	&= f_{ab}{^c}T_c + \Pi_{ab}^{ij}\,S\tfderr{}{\phi^j}\tfderl{}{\phi^i}\,,
	&
	\Pi^{ij}_{ab} &= (-)^{(\phi^i+1)(\phi^j+1)}\Pi^{ji}_{ab}\,,
\end{align}
where \(\Pi_{ab}^{ij}\) comprises the trivial symmetries \(\mu,\hat\mu,\hathat{\mu},\mu_+ \in \trivial\). The proper minimal solution to the global--gauge master equation is then given to second order in the antifields by
\begin{equation}
\label{bivector type BV solution}
\begin{split}
	S_\gBV
	&=
	S 
	+ \xi^a T_a{^i}\phia_i 
	- \tfrac12 (-)^{(\xi^a+1)\xi^b}\xi^a \xi^bf_{ab}{^c}\xi^+_c
	- \tfrac14 (-)^{(\xi^a+1)\xi^b + \phi^i\phi^j}\xi^a\xi^b\Pi_{ab}^{ij}\phia_i\phia_j
	\\
	&\eqcolon S + \xi^aT_a^\asta - Q_\CE^\asta + \Pi\,.
\end{split}
\end{equation}
When \eqref{bivector type BV solution} fully solves the global--gauge master equation, we say that the open Lie algebra representation is of the bivector type \cite[Def.~3.31]{Losev:2023gsq}. This essentially boils down to covariance of the \emph{bivector} under the symmetry group, as well as the Schouten--Nijenhuis bracket of the bivector with itself vanishing, encoded in respectively \cite{Losev:2023gsq}
\begin{align}
	(\xi^aT_a^\asta - Q_\CE^\asta,\Pi)_{\gBV} &= 0\,,
	&
	(\Pi,\Pi)_{\gBV} &= 0\,.
\end{align}

\paragraph{Closed Lie algebra representation with gauge symmetry.} We now include gauge symmetries into our discussion of solutions to the master equation. Let us denote the gauge symmetry generators by \(R_\alpha \in \Gamma(\tT\frF)\) and the global symmetry generators by \(T_a\in\Gamma(\tT\frF)\), with respective ghost fields \(c^\alpha\) and \(\xi^a\). These will generically obey relations
\begin{align}\label{Closed Lie algebra representation with gauge symmetry}
	[R_\alpha,R_\beta] &= f_{\alpha\beta}{^\gamma}R_\gamma
	\,,&
	[T_a,R_\beta] &= \rho(T_a)_\beta{^\gamma}R_\gamma
	\,,&
	[T_a,T_b] &= f_{ab}{^c}T_c + h_{ab}{^\gamma}R_\gamma
	\,.
\end{align}
Here, \(f_{\alpha\beta}{^\gamma}\), \(f_{ab}{^c}\) are respectively the structure constants of \(\frg\) and \(\frs\), and \(\rho(\phi)\) is a gauge-invariant representation of \(\frs\), and \(h_{ab}{^\gamma}(\phi)\) is both gauge- and \(\frs\)-invariant:
\begin{align}
    \label{structure tensor conditions}
	\big[\rho(T_a),\rho(T_b)\big]
	&= 
	f_{ab}{^c}\rho(T_c)\,,
	&
	R_\alpha\cdot\rho(T_a) &= 0\,,
	&
	T_a\cdot h_{ab}{^\gamma}(\phi) &= 0\,,
	&
	R_\alpha\cdot h_{ab}{^\gamma}(\phi) &= 0\,.
\end{align}
These two structure tensors measure the degree to which the representatives of \(T_a\) are not  gauge covariant. The gauge BV action, only taking into consideration gauge symmetries, is given by
\begin{equation}\label{eq:gauge BV action general case}
	S_\BV
	=
	S + c^\alpha R_\alpha^\asta - \tfrac12(-)^{(c^\alpha + 1)c^\beta}c^\alpha c^\beta f_{\alpha\beta}{^\gamma}c^+_\gamma\,.
\end{equation}
Extending now to include the global symmetries we obtain the global--gauge BV action
\begin{equation}
	S_\gBV = S_\BV 
	+ \xi^aT^\asta_a 
	- \tfrac12(-)^{(\xi^a+1)\xi^b}\xi^a\xi^b\Big(f_{ab}{^c}\xi^+_c + h_{ab}{^\gamma}(\phi)c^+_\gamma\Big)\,,
\end{equation}
where we defined
\begin{equation}
	T_a^\asta
	\coloneq 
	T_a{^i}\phi^+_i - (-)^{(\xi^a + 1)c^\beta}c^\beta\rho(T_a)_\beta{^\gamma}c^+_\gamma\,,
\end{equation}
extending the action of \(\frs\) to \(\frF_\gBV\).

\paragraph{Open Lie algebra representation of the bivector type with gauge symmetry.} Now, to move on to the case where the gauge symmetries close off-shell and the global symmetries only close on-shell, and are of the bivector type. The symmetry algebra then takes on the form
\begin{subequations}\label{eq:open Lie alg rep of bivec type w gauge symm}
\begin{align}
	[R_\alpha,R_\beta] &= f_{\alpha\beta}{^\gamma}R_\gamma
	\,,\Big.\\
	[T_a,R_\beta] &= \rho(T_a)_\beta{^\gamma}R_\gamma
	\,,\Big.\\
	[T_a,T_b] &= f_{ab}{^c}T_c + h_{ab}{^\gamma}R_\gamma + \Pi_{ab}^{ij}S\tfderr{}{\phi^j}\tfderl{}{\phi^i}
	\,.\Big.
\end{align}
\end{subequations}
Again, the structure tensors \(\rho(\phi)\) and \(h_{ab}{^\gamma}(\phi)\) satisfy the conditions \eqref{structure tensor conditions}. Furthermore, we require that the trivial symmetry \(\Pi_{ab} \in \trivial\) is gauge and globally invariant, and has a vanishing Schouten--Nijenhuis bracket with itself:
\begin{align}
	R_\alpha\cdot\Pi &= 0\,, &
	T_a\cdot\Pi &= 0\,, &
	[\Pi,\Pi] &= 0\,.
\end{align}
The global--gauge BV action then reads
\begin{equation}
	S_\gBV
	=
	S_\BV 
	+ \xi^aT^\asta_a 
	- \tfrac12(-)^{(\xi^a+1)\xi^b}\xi^a\xi^b\Big(f_{ab}{^c}\xi^+_c + h_{ab}{^\gamma}(\phi)c^+_\gamma\Big)
	+ \Pi\,,
\end{equation}
where \(\Pi\) as defined in \eqref{bivector type BV solution}.

\section{The superparticle and  Witten index}\label{sec:superparticle}

In this section, as a first example of the Batalin--Vilkovisky localisation scheme, we compute the Witten index \(\EV{1}\) for an \(\mathcal N=2\) superparticle living in a one-dimensional target space \(\bbR\) and subject to a superpotential \(h\).
Throughout this section, all of the computations will be performed on an \emph{on-shell}-closing representation for this superparticle. That is, the representation where the auxiliary field has been integrated out, and the algebra only closes on shell. However, for comparison with the on-shell case, we will also briefly review the off-shell-closing representation of the superparticle in this section. We will compute the Witten index in two ways: Firstly, we'll be applying the BV enhancement of standard localisation methods, painted in section \ref{sec:on shell localisation}, to obtain a localisation for the on-shell supermultiplet. Then, we'll be exploring a different approach to localisation, in which we mimmick standard gauge-fixing methods. Namely, we introduce trivial pairs, and achieve localisation by taking an \(R_\xi\)-gauge to a \(\delta\)-function gauge. These results will be shown to agree with results in the literature \cite{Witten:1982im}.
While the off-shell algebra is readily at hand for the superparticle (so the BV re-articulation  merely  illustrates the mechanism),  there are many  examples where this is much harder to realise. In such cases, the BV formulation may provide an alternative route.

We work on a periodic Euclidean worldline \(S^1\) of periodicity \(1/(2\pi T)\); the worldline coordinate will be denoted \(\tau\sim\tau + 1/(2\pi T)\). We take the superpotential \(h \in C^\infty(\bbR)\) to be a Morse function with finitely many extrema. Worldline derivatives will be denoted \(\dot{(-)}=\mathrm d/\mathrm d\tau\) while target-space derivatives will be denoted \((-)'=\mathrm d/\mathrm dx\). To avoid confusion, we will use \(\partial/\partial t\) to denote derivatives with respect to the localisation parameter \(t\).

\subsection{The \(d=1\), \(\cN = 2\) superparticle}

We start off with a brief review of the off-shell and on-shell formulations of the \(\cN = 2\) superparticle in one spacial dimension.

\paragraph{Off-shell formulation.} To realise \(d=1\), \(\mathcal N=2\) supersymmetry off shell, we take as the space of histories \(\frF^\off = C^\infty(S^1,\bbR^{2|2})\), corresponding to a bosonic scalar field \(x(\tau)\), two fermionic fields \(\psi(\tau)\) and \(\hat\psi(\tau)\), and an auxiliary scalar field \(F(\tau)\). The action \(S^\off \in C^\infty(\frF^\off)\) for the off-shell superparticle is given by
\begin{equation}\label{off shell superparticle action}
	S^\off
	=
	\int_{S^1} \extd\tau ~ \Big[\tfrac12 \dot{x}^2 + \tfrac12 h'(x)^2 + \hat\psi\big(\tder{}{\tau} - h''(x)\big)\psi + \tfrac12 F^2\Big]\,.
\end{equation}
This system has two supersymmetries \(\cQ,\hat\cQ \in \Gamma(\tT\frF^\off)\), which read
\begin{subequations}
\begin{align}
	\cQ &= \int_{S^1}\extd\tau ~ \Big[\psi\fder{}{x} + \big({-}\dot{x} + h' + \mathrm iF\big)\fder{}{\hat\psi} - \mathrm i\left(\der{}{\tau} - h''\right)\psi\fder{}{F}\Big]\,,
	\\
	\hat\cQ &= \int_{S^1}\extd\tau ~ \Big[{-}\hat\psi\fder{}{x} + \big(\dot x + h' + \mathrm iF\big)\fder{}{\psi} - \mathrm i\left(\der{}{\tau} + h''\right)\hat\psi\fder{}{F}\Big]\,.
\end{align}
\end{subequations}
These supersymmetries close into a \((2|1)\)-dimensional supersymmetry algebra \(\frs\) on \(\frF^\off\), 
\begin{align}
	[\cQ,\cQ] &= 0\,,
	&
	[\hat\cQ,\hat\cQ] &= 0\,,
	&
	[\cQ,\hat\cQ] &= 2\der{}\tau\,,
\end{align}
where \(\extd/\extd\tau = \int\extd\tau ~ (\dot x\delta/\delta x + \dot\psi\delta/\delta\psi + \cdots) \in \Gamma(\tT\frF^\off)\) the generator of Euclidean time translations. 
Now we extend the action \eqref{off shell superparticle action} to include global ghosts for the symmetry algebra \(\frs\). Using \eqref{off-shell BV solution}, the global Batalin--Vilkovisky action is given by
\begin{equation}\label{off shell ghobal superparticle action}
	S^\off_\gBV 
	=
	S^\off + \varepsilon\cQ^\asta + \hat\varepsilon\hat\cQ^\asta + \xi\cB^\asta - \varepsilon\hat\varepsilon\xia\,,
\end{equation}
where \(\varepsilon\), \(\hat\varepsilon\) and \(\xi\) are the global ghosts associated with respectively \(\cQ\), \(\hat\cQ\) and \(\cB = \der{}{\tau}\). These global ghosts are constants, rather than fields, in that they do \emph{not} depend on the Euclidean time coordinate. The final term corresponds to the Chevalley-Eilenberg differential term \(Q_\CE^\#\) in \eqref{bivector type BV solution}.

\paragraph{On-shell formulation.} The off-shell superparticle action \eqref{off shell superparticle action} contains an auxiliary field \(F\), which we may integrate out using its equation of motion \(F = 0\) to obtain the physically equivalent on-shell formulation. The space of histories is then \(\frF^\on = C^\infty(S^1,\bbR^{1|2})\) coordinatised by \(x(\tau)\), \(\psi(\tau)\), and \(\hat\psi(\tau)\), with the action
\begin{equation}\label{on shell superparticle action}
	S
	=
	S^\on
	= 
	\int_{S^1}\extd\tau ~ \Big[\tfrac12 \dot{x}^2 + \tfrac12 (h')^2 + \hat\psi\big(\tder{}{\tau} - h''\big)\psi\Big]\,.
\end{equation}
The supersymmetries \(\cQ,\hat{\cQ} \in \Gamma(\tT\frF^\on)\) are
\begin{align}
	\cQ &= \int_{S^1}\extd\tau ~ \Big[\psi\fder{}{x} + \big({-}\dot{x} + h'\big)\fder{}{\hat\psi}\Big]\,,
	&
	\hat\cQ &= \int_{S^1}\extd\tau ~ \Big[{-}\hat\psi\fder{}{x} + \big(\dot x + h'\big)\fder{}{\psi}\Big]\,.
\end{align}
They form an open (i.e.\ on-shell-closing) representation of the supersymmetry algebra since their commutators are
\begin{subequations}
\label{on-shell superparticle algebra}
\begin{align}
	[\cQ,\cQ] &= 0\phantom{\tder{}{\tau}} + \mu\,,
	&
	\mu &= S^\on\int_{S^1}\extd\tau ~ 2\fderr{}{\hat\psi}\fderl{}{\hat\psi}
	\\
	[\hat\cQ,\hat\cQ] &= 0\phantom{\tder{}{\tau}} + \hathat{\mu}\,,
	&
	\hathat{\mu} &= S^\on\int_{S^1}\extd\tau ~ 2\fderr{}{\psi}\fderl{}{\psi}
	\\
	[\cQ,\hat\cQ] &= 2\tder{}{\tau} + \hat\mu\,,
	&
	\hat\mu &= S^\on \int_{S^1}\extd\tau ~ \fderr{}{\psi}\fderl{}{\hat\psi} + \fderr{}{\hat\psi}\fderl{}{\psi}\,,
\end{align}
\end{subequations}
where \(\mu,\hat\mu,\hathat{\mu} \in \trivial\) are trivial symmetry transformations.
Extending the action \eqref{on shell superparticle action} to include global ghosts for the symmetry algebra \(\frs\) using \eqref{bivector type BV solution}, we obtain the Batalin--Vilkovisky action
\begin{equation}\label{on shell ghobal superparticle action}
	S_\gBV
	=
	S + \varepsilon \cQ^\asta + \hat\varepsilon\hat\cQ^\asta + \xi\cB^\asta - \varepsilon\hat\varepsilon\xia + \int_{S^1}\extd\tau ~ \tfrac12\big(\varepsilon\hat\psi^+ + \hat\varepsilon\psi^+\big)^2\,.
\end{equation}
The final term here corresponds to the bivector term \(\Pi\) in \eqref{bivector type BV solution}. The degree-\(0\) cohomology of \(Q_\gBV^{\off/\on}\) is evidently given by the BPS observables \(\cO_\BPS\) which are on-shell invariant under the whole supersymmetry algebra \(\frs\), i.e.\ \(\cQ\cO_\BPS \approx \hat\cQ\cO_\BPS \approx 0\).

\subsection{Localising subalgebras and equivariance conditions}
\label{ssec:Localising subalgebras and equivariance conditions}

We illustrate the systematics of BV enhanced localisation through two localising subalgebras.
\begin{itemize}
	\item Firstly, consider the \(0|1\)-dimensional \emph{nilpotent subalgebra} generated by a single supersymmetry \(\cQ\), owing its name to the fact that \([\cQ,\cQ] \approx 0\) (being a strict equality for the off-shell multiplet).
	\item Secondly, consider the \(1|1\)-dimensional \emph{equivariant subalgebra} generated by the diagonal combination \(\cQ_+ \coloneq \tfrac1{\sqrt2}\big(\cQ + \hat\cQ\big)\), and the corresponing time translation \(\cB \coloneq \tder{}{\tau} \approx \frac12[\cQ_+,\cQ_+]\) (again a strict equality for the off-shell multiplet).
\end{itemize}
In terms of global BV actions, the nilpotent case just amounts to turning off \(\hat\varepsilon\), \(\xi\) and their corresponding antifields. The equivariant case amounts to rotating the global ghosts to \(\varepsilon_\pm \coloneq \frac1{\sqrt2}(\varepsilon\pm\hat\varepsilon)\), and subsequently turning off \(\varepsilon_-\) and its antifield.

Suppose further that we consider an observable \(\cO_\BPS\) which is \emph{identically} invariant under the generators of the localising subalgebra in condsideration. That is, \(\cQ\cO_\BPS = 0\) for the nilpotent case and \(\cQ_+\cO_\BPS = \cB\cO_\BPS = 0\) for the equivariant case. Unpacking the BV--BPS equivariance conditions \eqref{BV BPS equivariance conditions} then yield \emph{no} equivariance conditions for the nilpotent subalgebra, and \(\der{}\tau\Psi(t) = 0\) for the equivariant subalgebra, the latter of which simply means that we take the localising fermion to be an integrated density on the Euclidean timeline, \(\Psi(t) = \int\extd\tau~(\cdots)\). This is all in spite of the fact that for the on-shell multiplets, \([\cQ,\cQ] = \mu\) and \([\cQ_+,\cQ_+] = \cB + \frac12(\mu + 2\hat\mu + \hathat\mu)\), which would ordinarily obstruct off-shell localisation arguments because of the trivial transformations.

\subsection{Nilpotent localisation scheme for the on-shell superparticle}
\label{ssec:Nilpotent localisation scheme for the on-shell superparticle}

We now move on to perform the localisation computation. We wish to perform this localisation using the nilpotent superalgebra generated by \(\cQ\). Following section \ref{ssec:Localising subalgebras and equivariance conditions}, we set \(\hat\varepsilon = \xi = 0\) (alongside their antifields) in action \eqref{on shell ghobal superparticle action} to obtain
\begin{equation}
	S_{\gBV} = S^\on + \varepsilon\cQ^\# + \varepsilon^2\int_{S^1}\extd\tau ~ \tfrac12(\hat\psi^+)^2\,.
\end{equation}
To construct the localising fermion, we now ring-theoretically localise the ring of functionals on global BRST configuration space to include a multiplicative inverse \(\frac1\varepsilon\) of the global ghost \(\varepsilon\), that is, \(C^\infty_\bullet(\frF_\gBRST^\times) = C^\infty(\frF)[\varepsilon,\frac1\varepsilon]\) (compare to section \ref{ssec:BRST reformulation of the off-shell localisation argument}).
The localising fermion \(\Psi(t) = \frac1\varepsilon\Psi_\loc(t)\) is not constrained by equivariance conditions, and we take it to be
\begin{equation}
	\Psi_\loc(t) 
	= \int_{S^1}\extd\tau ~ t\cdot\hat\psi(\cQ\hat\psi)
	\coloneq 
	\int_{S^1}\extd\tau ~ t\cdot\hat\psi(-\dot x + h')\,.
\end{equation}
Since the on-shell supersymmetry is manifestly real, there is no need to invoke an involution \((-)^\dagger : C^\infty(\frF)\to C^\infty(\frF)\) on the ring of complex functionals. The global BV action \(S_\gBV\) then pulls back to the global BRST action \(S_\gBRST\) on the Lagrangian submanifold \(\cL_{\Psi(t)}\), given by
\begin{equation}
\label{eq:deformed action superparticle}
\begin{split}
	S_\gBRST (t)
	\coloneq S_{\gBV}\big|_{\cL_{\Psi(t)}}
	&=
	S^\on + \cQ\Psi_\loc + \int_{S^1}\extd\tau ~ \tfrac12 \big(\tfder{}{\hat\psi}\Psi_\loc\big)^2
	\\
	&= \int_{S^1}\extd\tau ~ \Big[\tfrac12 (1+t)^2\big(\dot x^2 + (h')^2\big) + (1+t)\hat\psi\big(\tder{}{\tau} - h''\big)\psi\Big]\,.
\end{split}
\end{equation}
In the limit \(t\to\infty\), it is clear that the contributions localise to the locus \(\frF_\loc\) given by
\begin{align}
\label{eq:superparticle localisation locus}
	\dot x &= 0\,,
	&
	h' &= 0\,,
	&
	\psi &= \hat\psi = 0\,.
\end{align}
To perform the localisation, we expand perpendicularly to the localisation locus \(\tN\frF_\loc\) as
\begin{subequations}
\label{eq:expansion around localisation locus superparticle}
\begin{align}
	&
	\begin{aligned}
		x &= x_0 + \delta x
		\\
		&\coloneq x_0 + \tfrac1{1+t}\bar\delta x
	\end{aligned}\,,
	&
	&
	\begin{aligned}
		\psi &= 0 + \delta \psi
		\\
		&\coloneq 0 + \bar\delta\psi
	\end{aligned}\,,
	&
	&
	\begin{aligned}
		\hat\psi &= 0 + \delta \hat\psi
		\\
		&\coloneq 0 + \tfrac1{1+t}\bar\delta\hat\psi
	\end{aligned}\,.
\end{align}
Note that the rescaling weights aren't uniform across all field components. This reflects the fact that we work with an on-shell multiplet in which the off-shell bosonic--fermionic degrees of freedom aren't matched one-to-one (compare to \eqref{eq:theory rescaling mixed approach} for the case of an off-shell-closing multiplet).
This is further reflected in the Jacobian factor, which reads
\begin{equation}
	\Ber\fder{(\delta x,\delta\psi,\delta\hat\psi)}{(\bar\delta x,\bar\delta\psi,\bar\delta\hat\psi)}
	=
	\frac{\Det 1\cdot\Det(1+t)}{\Det(1+t)} = 1\,,
\end{equation}
where we used \(\zeta\)-function regulsarisation to define the functional determinant \(\Det(1+t)^p = (1+t)^{p\zeta(0)} = (1+t)^{-p/2}\), with \(\zeta(z)\) the Riemann \(\zeta\)-function, which regulates \(1+1+\cdots \overset{\text{reg}}{=} \zeta(0) = -\frac12\).
Indeed, were we to use uniform rescaling weights, the overall Jacobian factor would be a non-vanishing power of \(1+t\), spoiling the localisation argument.

Expanding the action in the normal directions around the localisation locus at a stationary point \(x_0\) of the superpotential, we find that the action is global BRST action \(S_\gBRST\) becomes dominated by the free terms, as \(t\) goes to infinity,
\begin{equation}
\begin{aligned}
	S_\gBRST(t)
	&=
	S_\free + O(\tfrac1{1+t})\,,
	\\
	S_\free &= \int_{S^1}\extd\tau ~ \Big[\tfrac12\bar\delta\dot x^2 + \tfrac12 h''(x_0)^2\bar\delta x^2 + \bar\delta\hat\psi\big(\tder{}{\tau} - h''(x_0)\big)\bar\delta\psi\Big]\,.
\end{aligned}
\end{equation}
\end{subequations}
Since we can compute path integrals of free theories, we can use the \(t\to\infty\) limit to evaluate the path integral expression for the Witten index,
\begin{equation}
\begin{split}
	\EV1
	=
	\lim_{t\to\infty}\EV1(t)
	&\overeq{\ref{eq:superparticle localisation locus}}
	\lim_{t\to\infty}\sum_{x_0}\int_{\tN_{x_0}\frF}\extD\delta x\,\extD\delta\psi\,\extD\delta\hat\psi ~ \rme^{-S_\gBRST(t)}
	\\
	&\overeq{\ref{eq:expansion around localisation locus superparticle}}
	\sum_{x_0}\int_{\tN_{x_0}\frF}\extD\bar\delta x\,\extD\bar\delta\psi\,\extD\bar\delta\hat\psi ~ \rme^{-S_\free}
	\\
	&= \sum_{x_0}\frac{1}{\Ber^{\frac12}\Hess S_\free\big|_{x_0}}\,,
\end{split}
\end{equation}
where the Berezinian factors are evaluated to
\begin{equation}
\begin{split}
	\Ber^{\frac12}\Hess S_\free \big|_{x_0}
	&= 
	\Det^{\frac12}\big(-\tder{^2}{\tau^2} + h''(x_0)\big)\cdot\Det^{-1}\big(\tder{}{\tau} - h''(X_0)\big)
	\Big.\\
	&= \big|\sinh\big(\tfrac1{4\pi T}h''(x_0)\big)\big|\cdot\sinh\big(\tfrac1{4\pi T}h''(x_0)\big)^{-1}
	\Big.\\
	&= \sign h''(x_0)\,.
	\Big.
\end{split}
\end{equation}
Thus, our computation yields an expression for the Witten index given by
\begin{equation}
	\EV1 = \sum_{x_0}\sign h''(x_0)\,.
\end{equation}
This agrees with standard results in the literature \cite{Witten:1982im}.

In summary, we started off with an on-shell-closing representation \((x,\psi,\hat\psi)\) for an \(\cN = 2\) superparticle in one spacial dimension. By recasting the localisation argument in the BV formalism, we managed to circumvent the fact that the supersymmetry \(\cQ\) squares into a trivial symmetry \([\cQ,\cQ] = \mu\), which in the off-shell framework would lead to the problematic equivariance condition \(\mu\Psi_\loc = 0\). Instead, in the BV localisation framework, this issue is circumvented because the localising fermion is unconstrained. A new phenomenon in the computation is the fact that the rescalings \eqref{eq:expansion around localisation locus superparticle} of the fluctuations around the localisation locus are now non-uniformly weighted, so as to accommodate the fact that the off-shell bosonic--fermionic degrees of freedom cease to be one-to-one.

\subsection{Localisation as an \(R_\xi\)-gauge}
\label{ssec:R xi gauge localisation superparticle}

Motivated by the perspective on localisation as gauge-fixing, we now bring to light a different perspective on localisation. We aim to pursue an approach to localisation which mirrors imposing an \(R_\xi\)-gauge (where \(\xi\) not to be confused with the global symmetry ghost). In particular, as \(\xi\to 0\), an \(R_\xi\)-gauge becomes a \(\delta\)-function gauge, concentrated in the gauge-fixing condition. This will be mirrored in the context of localisation where taking \(t\to\infty\) corresponds to a \(\delta\)-function gauge concentrated in the localisation locus. 

To this end, we start off by introducing two trivial pairs \((\sigma(\tau),\beta(\tau))\) and \((\sigmat(\tau),\betat(\tau))\) of respective bidegrees \(|\sigma| = (-1,\even)\), \(|\beta| = (0,\even)\), \(|\sigmat| = (-1,\odd)\) and \(|\betat| = (0,\odd)\). These contribute to the global BV action as\footnote{The factor `\(\rmi\)' in \eqref{eq:extension of global BV action superparticle} arises from our conventions for the path integral representations of functional \(\delta\)-functions. We work in the convention \(\delta[f] = \int\extD\bos ~ \exp\rmi\int\extd\tau ~ \bos(\tau)f(\tau)\) for \(f\) a bosonic function and \(\delta[\vartheta] = \int\extD\fer ~ \exp\int\extd\tau ~ \fer(\tau)\vartheta(\tau)\) for \(\vartheta\) a fermionic function.}
\begin{equation}
\label{eq:extension of global BV action superparticle}
	S_\gBV\big|_\old
	\to
	S_\gBV\big|_\new \coloneq S_\gBV\big|_\old + \int_{S^1}\extd\tau ~ \Big[\rmi\beta\sigma^+ + \betat\sigmat^+\Big]\,.
\end{equation}
Consider now the localisation fermion
\begin{equation}
	\Psi(t)
	=
	\int_{S^1}\extd\tau ~ \Big[
	-\varepsilon\sigma\sigmat + t\hat\psi\sigmat + \sigma\cQ\hat\psi
	\Big]\,.
\end{equation}
This construction mirrors the gauge-fixing fermion \eqref{eq:R xi gauge fixing fermion} for an \(R_\xi\)-gauge, in that the `global antighosts' \(\sigma\) and \(\sigmat\) pair up with respective `global gauge fixing conditions' \(t\cQ\hat\psi\) and \(\hat\psi\), and there is a quadratic term.
This localising fermion gives rise to a global BRST action
\begin{equation}
	S_\gBRST(t)
	=
	S
	+
	\int_{S^1}\extd\tau ~ \Big[
	\begin{aligned}[t]
	&\rmi\beta\big(\cQ\hat\psi - \varepsilon\sigmat\big) + \betat\big(t\hat\psi - \varepsilon\sigma\big)
	\\
	&- (\varepsilon\sigma)\tfrac12\mu\hat\psi + t(\varepsilon\sigmat)\cQ\hat\psi + \tfrac12 t^2(\varepsilon\sigmat)^2\Big]
	\end{aligned}
\end{equation}
One now sees that the fields \(\sigma\), \(\beta\) and \(\betat\) becomes Lagrange multipliers, whereas the field \(\sigmat\) becomes an auxiliary field. The reason that the field \(\sigmat\) is not also a Lagrange multiplier is that we're working with an on-shell-closing multiplet, and this compensates for the failure of one-to-one matching of off-shell bosonic--fermionic degrees of freedom.
Furthermore, integrating out the Nakanishi--Lautrup fields \(\beta\) and \(\betat\) at arbitrary \(t\geq 0\), one recovers the global BRST action \eqref{eq:deformed action superparticle}. For this operation to be well-defined, we again ring-theoretically localised the space of functionals to include a multiplicative inverse \(\frac1\varepsilon\) of the global ghost \(\varepsilon\)\footnote{More specifically, suppose \(\varepsilon\) doesn't have a multiplicative inverse, then \(\delta\)-functionals \(\delta[\cQ\hat\psi - \varepsilon\sigmat]\) and \(\delta[t\hat\psi - \varepsilon\sigma]\) obtained by integrating out the Nakanishi--Lautrup fields \(\beta\) and \(\betat\) cannot be used to substitute \(\varepsilon\sigmat = \cQ\hat\psi\) and \(\varepsilon\sigma = t\hat\psi\) without also constraining the other field content.}.
At \(t = 0\), this establishes that using such a localising fermion does not change the physics, and for \(t > 0\) this shows the equivalence between this approach to localisation and the previous.

To make manifest the role of \(t\) as the parameter of an \(R_\xi\)-gauge, we need to perform a rescaling at \(t>0\), given by
\begin{align}
	&
\begin{aligned}
	\sigmat &\to \tfrac1t\sigmat\,,
	\\
	\betat &\to \tfrac1t\betat\,,
\end{aligned}
	&
	&
	\begin{aligned}
	S_\gBRST(t) \to S + 
	\int_{S^1}\extd\tau ~ \Big[
	\begin{aligned}[t]
	&\rmi\beta\big(\cQ\hat\psi - \tfrac1t\varepsilon\sigmat\big) + \betat\big(\hat\psi - \tfrac1t\varepsilon\sigma\big)
	\\
	&- (\varepsilon\sigma)\tfrac12\mu\hat\psi + (\varepsilon\sigmat)\cQ\hat\psi + \tfrac12 (\varepsilon\sigmat)^2\Big]\,.
	\end{aligned}
	\end{aligned}
\end{align}
(This rescaling clearly has a unit Jacobian factor.) Indeed, now the terms quadratic in the trivial pairs (aside from the auxiliary field \(\sigmat\)) appear with a factor \(t^{-1}\), playing the role of \(\xi\) in the \(R_\xi\)-gauge. Taking \(\xi\to 0\) to impose a \(\delta\)-function gauge then corresponds to taking \(t\to\infty\). In this limit the global BRST action becomes
\begin{equation}
\label{eq:R xi gauge action superparticle after rescaling}
	S_\gBRST(t)
	\xrightarrow{t\to\infty}
	S
	+
	\int_{S^1}\extd\tau ~ \Big[
	\rmi\beta\cQ\hat\psi
	+ \betat\hat\psi
	- (\varepsilon\sigma)\tfrac12\mu\hat\psi
	+ (\varepsilon\sigmat)\cQ\hat\psi + \tfrac12 (\varepsilon\sigmat)^2
	\Big]\,.
\end{equation}
We now proceed to integrate out the trivial pairs, serving as Lagrange multipliers and auxiliary fields, from the expression of the Witten index \(\EV1\). To this end, we'd like to treat \(\varepsilon\sigma\) and \(\varepsilon\sigmat\) as degree-\(0\) fields which we integrate over. 
The transformations \(\sigma\to\varepsilon\sigma\) and \(\sigmat\to\varepsilon\sigmat\) necessitate ring-theoretically localising the space of functionals to include a multiplicative inverse \(\frac1\varepsilon\) of the global ghost \(\varepsilon\) again for this transformation to be invertible.
We can then rewrite the measure as \(\extD\sigma\,\extD\sigmat = \extD(\varepsilon\sigma)\,\extD(\varepsilon\sigmat)\), since \(\sigma\) and \(\sigmat\) carry opposite fermion parity. Thus, integrating out the trivial pairs we obtain
\begin{equation}
\label{eq:superparticle computation}
\begin{split}
	\EV1 
	&= \lim_{t\to\infty}\int\extD x\,\extD\psi\,\extD\hat\psi\,\extD(\varepsilon\sigma)\,\extD\beta\,\extD(\varepsilon\sigmat)\,\extD\betat ~ \rme^{-S_\gBRST(t)}
	\\
	&\overeq{\ref{eq:R xi gauge action superparticle after rescaling}} \int\extD x\,\extD\psi\,\extD\hat\psi ~ \delta\big[\cQ\hat\psi\big]\delta\big[\hat\psi\big]\delta\big[{-}\tfrac12\mu\hat\psi\big]\,.
\end{split}
\end{equation}
The three functional \(\delta\)-functions come respectively from integrating out the fields \(\beta\), \(\betat\) and \(\sigma\), and the functional determinant factor comes from integrating out \(\sigmat\). Let us now make sense of this expression by using common \(\delta\)-function identities. Particularly, we will make use of the identities
\begin{align}
\label{eq:properties delta functions}
	\delta_\bos(f(x)) &= \sum_{x_0 \in f^{-1}(0)} \frac{\delta_\bos(x-x_0)}{|f'(x_0)|}\,,
	&
	\delta_\fer(a\theta) &= a\delta_\fer(\theta)\,.
\end{align}
Accordingly, we can reduce two of the \(\delta\)-functions to
\begin{equation}
\label{eq:computation delta functions superparticle}
\begin{aligned}
	\delta\big[\cQ\hat\psi\big]
	&=
	\delta\big[-\dot x + h'\big]
	=
	\sum_{x_0 \in h^{-1}(0)}\frac{\delta\big[x - x_0\big]}{\big|\Det\big(\tder{}{\tau} - h''(x_0)\big)\big|}
	\\
	\delta\big[{-}\tfrac12\mu\hat\psi\big]
	&=
	\delta\big[\big(\tder{}{\tau} - h''\big)\psi\big]
	=
	\Det\big(\tder{}{\tau} - h''\big)\cdot\delta\big[\psi\big]
\end{aligned}
\end{equation}
Substituting these in the path integral \eqref{eq:superparticle computation} we find
\begin{equation}
\begin{split}
	\EV1
	&\overeq{\ref{eq:computation delta functions superparticle}}
	\int\extD x\,\extD\psi\,\extD\hat\psi ~ \sum_{x_0}
	\frac{\Det\big(\tder{}{\tau} - h''\big)}{\big|\Det\big(\tder{}{\tau} - h''(x_0)\big)\big|}\delta\big[x-x_0\big]\delta\big[\psi\big]\delta\big[\hat\psi\big]
	\bigg.\\
	&= \sum_{x_0}\frac{\Det\big(\tder{}{\tau} - h''(x_0)\big)}{\big|\Det\big(\tder{}{\tau} - h''(x_0)\big)\big|}
	\bigg.\\
	&= \sum_{x_0}\sign h''(x_0)
	\bigg.\,,
\end{split}
\end{equation}
which is in agreement with the results in section \ref{ssec:Nilpotent localisation scheme for the on-shell superparticle} and the literature \cite{Witten:1982im}.

In summary, motivated by the perspective on localisation as gauge-fixing, we proceeded to reformulate the BV localisation procedure as a gauge-fixing procedure. To this end, we introduced two trivial pairs, and wrote down a localising fermion, which to an extent mirrors the stucture of a standard localising fermion for an \(R_\xi\)-gauge. In this context, \(t^{-1}\) takes on the role of the parameter \(\xi\), and much like how \(\xi\to 0\) produces a \(\delta\)-function gauge, \(t^{-1}\to 0\) produces the localisation.

\newcommand{\zetad}{\zeta^\dagger}
\newcommand{\Qt}{\accentset{\sim}{Q}}
\newcommand{\extDs}{\slashed{\extD}}
\newcommand{\Fs}{\slashed{F}}
\newcommand{\sfL}{\mathsf{L}}
\newcommand{\eb}{\bar{e}}
\newcommand{\sfc}{\mathsf{c}}
\newcommand{\zetac}{\zeta^\sfc}
\newcommand{\zetacd}{\zeta^{\sfc\dagger}}
\newcommand{\muhh}{\hathat{\mu}}
\newcommand{\muh}{\hat{\mu}}
\newcommand{\cc}{\mathsf{c}}
\newcommand{\extDsr}{\accentset{\leftarrow}{\extDs}}
\newcommand{\adr}{\accentset{\leftarrow}{\ad}}
\newcommand{\Ks}{\slashed{K}}
\newcommand{\Ab}{\bar{A}}
\newcommand{\sigmab}{\bar{\sigma}}
\newcommand{\Db}{\bar{D}}
\newcommand{\lambdatb}{\bar{\lambdat}}
\newcommand{\tU}{\text{U}}

\section{\(d = 3\), \(\mathcal N = 2\) supersymmetric Yang--Mills theory on Seifert manifolds}\label{sec:seifert}

In this section, in studying \(d = 3\), \(\mathcal N = 2\) supersymmetric Yang--Mills theory, we consider a first case for BV localisation applied to a theory with non-trivial gauge symmetries. The overall structure of this section will be the same as that of the previous section.
The traditional discussion of localisation of \(d=3\), \(\mathcal N=2\) theories requires off-shell supersymmetry with auxiliary fields \cite{Kallen:2011ny,Beasley:2005vf,Closset:2012ru,Dumitrescu:2016ltq,Willett:2016adv}; we show that the auxiliary fields may be dispensed with in the Batalin--Vilkovisky formalism.
For the sake of brevity we will restrict ourselves to the nilpotent localisation scheme for the on-shell multiplet. 

\subsection{Lightning review of Seifert manifolds}
We work with \emph{Seifert manifolds}, which we define as the total spaces of  principal \(\operatorname U(1)\)-bundles on a Riemann surface \(\Sigma\) (see reviews \cite{lee2001seifertmanifolds,brin2007seifertfiberedspacesnotes}),
\begin{equation}
	\operatorname U(1)\to M \onto \Sigma \coloneq M/\operatorname U(1)\,.
\end{equation}
We take these to be equipped with a Riemannian metric \(g\) (Hodge star denoted \(\star\)) such that the \(\operatorname U(1)\)-action on \(M\) is an isometry.
We further assume that \(M\) is a spin manifold, with spinor bundle \(S\onto M\) such that there exists a Killing spinor \(\zeta \in \Gamma(S)\) satisfying the Killing spinor equation
\begin{equation}\label{eq:Killing spinor equation 3d}
    \nabla\zeta = \frac\rmi2H\gamma\zeta
\end{equation}
for some real number \(H > 0\). We take this Killing spinor to define the generator \(K \in \Gamma(\tT M)\) of the \(\operatorname U(1)\)-action and the contact structure one-form \(\kappa \in \Omega^1(M)\) on \(M\),
\begin{align}\label{eq:seifert contact structure}
	K^\mu &= \frac{1}{\zeta^\dagger\zeta}\zeta^\dagger\gamma^\mu\zeta\,,
	&
	\kappa &= \frac{1}{\zeta^\dagger\zeta}\zeta^\dagger\gamma\zeta\,.
\end{align}
Applying some spinor calculus (expanded on in appendix \ref{Appendix}) and the Killing spinor equation \eqref{eq:Killing spinor equation 3d} one can show the following identities:
\begin{align}
	\extd(\zeta^\dagger\zeta) &= 0\,&
	\extd\kappa &= 2H\star\kappa\,,&
	K\inp \kappa &= 1\,,
\end{align}
It thus follows that \(K\) is the Reeb vector field for the contact structure \(\kappa\).
A stadard example is the Hopf fibration, \(S^1 \into S^3 \onto S^2\), for which \(H\) is the inverse of the radius of the round 3-sphere \(S^3\).
More useful identities can be found in appendix~\ref{Appendix}.

The \(\mathcal N = 2\) supersymmetry algebra we consider below is parameterised by the Killing spinor \(\zeta \in \Gamma(S)\) satisfying the Killing spinor equation \eqref{eq:Killing spinor equation 3d}, chosen such that the Reeb vector field \(K\) of the corresponding contact structure defines a \(\operatorname U(1)\)-fibration.

\subsection{The global--gauge Batalin--Vilkovisky action}

\paragraph{Gauge BV action.} The field content \(\frF_P\) of the on-shell multiplet of \(\cN = 2\) supersymmetric Yang--Mills theory on a Seifert manifold \(M\) with gauge group \(G\) consists, for a given principal \(G\)-bundle \(P\onto M\), of
\begin{itemize}
	\item a gauge connection \(A \in \cA(P)\),
	\item an adjoint scalar field \(\sigma \in \Gamma(\ad P)\), and
	\item two adjoint gauginos \(\lambda \in \Gamma(\Pi S\otimes\ad P)\) and \(\hat\lambda \in \Gamma(\Pi S^\ast\otimes\ad P)\),
\end{itemize}
where we defined the adjoint bundle \(\ad P \coloneq P\times_{\Ad} \frg\) to be the associated bundle for the adjoint representation on the underlying Lie algebra \(\frg \coloneq \Lie(G)\). To complete this to an off-shell multiplet, one would have to further add an auxiliary adjoint scalar \(D\in\Gamma(\ad P)\). However, for the purposes of demonstrating localisation in the Batalin--Vilkovisky formalism, we will focus on the on-shell multiplet.
Since the underlying principal bundle is also dynamical, we need to sum over the equivalence classes \([P]\) of principal bundles, \(\frF = \coprod_{[P]}\frF_P\). 
The action for the on-shell multiplet reads
\begin{equation}
	S
	=
	\frac1{g_\ym^2}\int_M \tfrac12 F_A\wedge\star F_A 
	+ \tfrac12\extD_A\sigma\wedge\star \extD_A\sigma 
	+ \star\hat\lambda\big(\rmi\extDs_A - \ad\sigma - \tfrac12 H\big)\lambda\,,
\end{equation}
where we defined the field strength \(F_A \coloneq \extd A + \tfrac12[A,A]\) and covariant derivative \(\extD_A \coloneq \nabla + \ad A\). Feynman slashes are normalised as \(\extDs_A = \gamma^\mu\extD_\mu\) and \(\Fs = \tfrac12\gamma^{\mu\nu}F_{\mu\nu}\). We further also take \(\frg\) to be a quadratic Lie algebra and implicitly contract with a \(G\)-invariant positive-definite form \(\EV{-,-}_\frg : \frg\otimes\frg\to\bbR\). The action for the off-shell multiplet is obtained by adding a \(g_\ym^{-2}\int\star\tfrac12 (D-H\sigma)^2\) term.

The finite gauge symmetries \(\cG_P\) on a given component \(\frF_P\) of configuration space are given by the sections of the (Lie group) adjoint bundle, \(\cG_P = \Gamma(\Ad P)\), where we defined \(\Ad P \coloneq P\times_{\Ad}G\). Accordingly, the infinitesimal gauge symmetries correspond to sections of the (Lie algebra) adjoint bundle, \(\frg_P = \Gamma(\ad P)\). These follow ansatz \eqref{eq:gauge BV action general case}, and accordingly, the (minimal) gauge BV action \(S_\BV\in C^\infty(\frF_{P|\BV})\) is given by
\begin{equation}
	S_\BV
	=
	S + \int_M-\extD_Ac\wedge A^+ + [c,\sigma]\sigma^+ + \lambda^+[c,\lambda] + [c,\hat\lambda]\hat\lambda^+ + \tfrac12[c,c]c^+\,,
\end{equation}
where we introduced a ghost field \(c \in \Gamma(\ad P)[1]\). Here, antifields of \(p\)-form fields are \(p\)-vector density fields, which are naturally isomorphic to \((3-p)\)-form fields. It is straightforward to modify this action for the off-shell multiplet.

\paragraph{Gauge fixing.} We further note that for the purposes of gauge fixing we will add a trivial pair \((\bar c,b) \in \tT[1]\Gamma(\ad P)[-1]\), consisting of an antighost field \(\bar c \in \Gamma(\ad P)[-1]\) and a Nakanishi--Lautrup field \(b \in \Gamma(\ad P)\), extending the minimal gauge BV action to the extended gauge BV action,
\begin{equation}
	S_\BV|_\old \to S_\BV|_\new = S_\BV|_\old + \int_M\extd^3x ~ \rmi b\bar c^+\,.
\end{equation}
Throughout subsequent sections, we impose the Lorenz gauge as a \(\delta\)-function gauge through the gauge-fixing fermion
\begin{equation}
\label{eq:gf fermion Lorenz gauge}
	g_\ym^2\Psi_\gf = \int_M\star\bar c(\nabla\inp A)\,.
\end{equation}
Note that this gauge-fixing procedure suffers from the Gribov ambiguity, which we will neglect. For a systematic approach to dealing with this issue in the BV formalism, we refer the reader to \cite{Getzler:2019etm}. We will further also neglect subtleties involving ghost zero modes. This issue is addressed in the context of localisation in \cite[\S{}4.1]{Pestun:2007rz}.
Accordingly, the gauge BRST action is given by
\begin{equation}
	S_\BRST 
	\coloneq S_\BV\big|_{\cL_\gf}
	= S + \int_M\star \rmi b(\nabla\inp A) - \nabla\bar c\wedge\star\extD_Ac\,.
\end{equation}
The path integral expression for the expectation value of a gauge observable \(\cO\) thus reads
\begin{equation}
\label{eq:EV observable SYM}
	\EV\cO
	=
	\sum_{[P]}\int_{\frF_{P|\BRST}}\extD A\,\extD\sigma\,\extD\lambda\,\extD\hat\lambda\,\extD c\,\extD\bar c\,\extD b ~ \cO\rme^{-S_\BRST}\,,
\end{equation}
where we have to sum over equivalence classes \([P]\) of principal \(G\)-bundles \(P\onto M\).

\paragraph{Supersymmetry algebra.} The supersymmetry algebra is generated by two supersymmetries \(\cQ,\hat\cQ\in\Gamma(\tT\frF_P)\), which are respectively parametrised by the Killing spinor \(\zeta \in \Gamma(S)\) and its Hermitian conjugate \(\zetad \in \Gamma(S^\ast)\), and read
\begin{equation}
\begin{aligned}
	\cQ A_\mu &= -\zeta^\dagger\gamma_\mu\lambda\,,
	&
	\cQ\sigma &= -\rmi\zetad\lambda\,,
	&
	\cQ\lambda &= 0\,,
	&
	\cQ\hat\lambda &= \rmi\zetad\big(\Fs_A - \rmi\extDs_A\sigma\big)\,,
	\\
	\hat\cQ A_\mu &= \phantom{-^\dagger}\hat\lambda\gamma_\mu\zeta\,,
	&
	\hat\cQ\sigma &= \phantom{-^\dagger}\rmi\hat\lambda\zeta\,,
	&
	\hat\cQ\lambda &= \rmi\big(\Fs_A + \rmi\extDs_A\sigma\big)\zeta\,,
	&
	\hat\cQ\hat\lambda &= 0\,.
\end{aligned}
\end{equation}
To modify the supersymmetries \(\cQ = \cQ^\on\) for the off-shell multiplet supersymmetries \(\cQ^\off\) one performs changes
\begin{align}
	&
	\begin{aligned}
		\cQ^\off\hat\lambda &= \cQ^\on\hat\lambda + \rmi\zetad(D-H\sigma)\,,
		\\
		\hat\cQ^\off\lambda &= \hat\cQ^\on\lambda + \phantom{\zetad}\rmi(D-H\sigma)\zeta\,,
	\end{aligned}
	&
	&
	\begin{aligned}
		\cQ^\off D &= \extD_\mu(\zetad\gamma^\mu\lambda) + \rmi(\ad\sigma + H)\zetad\lambda\,,
		\\
		\cQ^\off D &= \extD_\mu(\hat\lambda\gamma^\mu\zeta)\phantom{^\dagger} + \rmi(\ad\sigma -H)\hat\lambda\zeta\,.
	\end{aligned}
\end{align}
However, much like section \ref{sec:superparticle}, we will mainly focus on the on-shell multiplet to showcase the novelties of localisation in the Batalin--Vilkovisky formalism.
On the on-shell multiplet, the supersymmetry transformations close into
\begin{subequations}
\begin{align}
	[\cQ,\cQ] &= \phantom{\rmi(\zetad\zeta)\cB}0 + g_\ym^2\mu\,,
	&
	\mu &= S\int_M\frac{\extd^3x}{\sqrt{\det g}} ~ 2\big(\zetad\fderr{}{\hat\lambda}\big)\big(\zetad\fderl{}{\hat\lambda}\big)\,,
	\\
	[\cQ,\hat\cQ] &= 2\rmi(\zetad\zeta)\cB + g_\ym^2\hat\mu\,,
	&
	\hat\mu &= S\int_M\frac{\extd^3x}{\sqrt{\det g}} ~ \big(\zetad\fderr{}{\hat\lambda}\big)\big(\fderl{}{\lambda}\zeta\big) + \big(\fderr{}{\hat\lambda}\zeta\big)\big(\zetad\fderl{}{\hat\lambda}\big)\,,
	\\
	[\hat\cQ,\hat\cQ] &= \phantom{\rmi(\zetad\zeta)\cB}0 + g_\ym^2\hathat\mu\,,
	&
	\hathat\mu &= S\int_M\frac{\extd^3x}{\sqrt{\det g}} ~ 2\big(\fderr{}{\lambda}\zeta\big)\big(\fderl{}{\lambda}\zeta\big)\,,
	\\
	[\cQ,\cB] &= [\hat\cQ,\cB] = [\cB,\cB] = 0\,,
	\bigg.
\end{align}
\end{subequations}
where \(\mu,\hat\mu,\hathat\mu\in\trivial\) are trivial symmetry transformations. The global symmetry \(\cB\) is given by covariantised rotation around the \(\tU(1)\)-fibre,
\begin{equation}
	\cB \coloneq \cL_K^\nabla + \delta_{\frg_P}(K\inp A + \rmi\sigma) - \delta_\sfR(\tfrac12 H) + \delta_\sfL\big(\tfrac12\nabla_aK_bM^{ab}\big)\,.
\end{equation}
Here, \(\cL_K^\nabla \coloneq [\nabla,\iota_K]\) is the Lorentz-covariantised Lie derivative, \(\delta_\frg(-)\) denotes a local gauge transformation, \(\delta_\sfR(-)\) denotes a \(\operatorname U(1)_\sfR\)-transformation, and \(\delta_\sfL(-)\) denotes a local Lorentz transformation. The \(\operatorname U(1)_\sfR\)-charges of the fields \((A,\sigma,\lambda,\hat\lambda,D)\) are \((0,0,+1,-1,0)\), and the Lorentz algebra generators \(M^{ab}\) are normalised to obey \([M_{ab},M^{cd}] = -4\delta_{[a}{^{[c}}M_{b]}{^{d]}}\), so that in particular its spin-\(\frac12\) representation is given by \(\rho(M^{ab}) = \frac12\gamma^{ab}\). When one modifies these transformations to the off-shell supersymmetries \(\cQ^\off\), the same algebra is obeyed off shell, so that the trivial symmetry transformations vanish, \(\mu^\off = \hat\mu^\off = \hathat\mu^\off = 0\).

For the off-shell multiplet, these transformations close on behalf of the \(\bbZ_2\)-graded Jacobi identity. Note, however, that large global symmetry transformations only close up to (field-dependent) gauge transformations, which is reflected by the fact that global symmetries act uniquely on gauge orbits, and that one needs to non-canonically lift a transformation on the gauge orbits to a transformation on field configurations. For the on-shell multiplet, it goes even further, since these only close on the critical locus. These showcase the different ways in which the symmetry algebra ceases to be a conventional representation of a group, as was discussed in section \ref{ssec:Blurring the distinction between global and gauge symmetries}.

\paragraph{Global--gauge BV action.} We note that the global symmetry structure is of the bivector type, as described in \eqref{eq:open Lie alg rep of bivec type w gauge symm}. Thus, introducing degree-\((+1,\odd)\) global ghosts \(\varepsilon,\hat\varepsilon\) for the supersymmetries \(\cQ,\hat\cQ\) and a degree-\((+1,\even)\) global ghost \(\xi\) for the isometry \(\cB\), alongside their respective antifields, the global--gauge BV action \(S_\gBV\) for the on-shell multiplet reads
\begin{equation}
	S_\gBV = S_\BV + \varepsilon\cQ^\# + \hat\varepsilon\hat\cQ^\# + \xi\cB^\# -2\rmi\varepsilon\hat\varepsilon\xi^+ + \int_M\frac{\extd^3x}{\sqrt{\det g}} ~ \tfrac12 g_\ym^2\big(\hat\varepsilon\lambda^+\zeta + \varepsilon\zetad\hat\lambda^+\big)^2\,.
\end{equation}
For the purposes of localisation, we will specialise to the \(0|1\)-dimensional nilpotent subalgebra generated by \(\cQ\), effectively setting \(\hat\varepsilon = \xi = 0\) alongside their antifields (compare to \cref{ssec:Localising subalgebras and equivariance conditions,ssec:Nilpotent localisation scheme for the on-shell superparticle}). This gives us a global--gauge BV action
\begin{equation}
\label{eq:global-gauge BV action SYM}
	S_\gBV \to S_\gBV = S_\BV + \varepsilon\cQ^\# + \varepsilon^2\int_M\frac{\extd^3x}{\sqrt{\det g}} ~ \tfrac12 g_\ym^2\big(\zetad\hat\lambda^+\big)^2\,.
\end{equation}
The symmetry algebra for off-shell multiplet is not of the bivector type, and accordingly, its global--gauge BV action will not include second order antifield terms. 
Henceforth we will consider a generic BPS observable \(\cO_\BPS\), taken to satisfy \(\cQ\cO_\BPS = 0\) identically.
We further note that the combined localising fermion \(\Psi(t)\) will not be constrained by any equivariance conditions in BV localisation schemes, following the BV--BPS equivariance condition \eqref{BV BPS equivariance conditions}, since neither the global--gauge BV action nor the BV--BPS observable \(\cO_\BPS\) depend on \(\varepsilon^+\).

\subsection{Batalin--Vilkovisky localisation}

Now, to perform localisation in the presence of gauge symmetries, we need to contend with the fact that the off-shell  modes normal to the localisation locus do not enjoy a boson--fermion one-to-one correspondence. There are two reasons for this:
\begin{itemize}
	\item Firstly, when considering off-shell-closing multiplets for gauge theories, bosonic modes which are `pure gauge' don't pair up with fermionic modes, since they correspond to redundancies in the description of the theory.
	\item Secondly, there is also a discrepancy between bosonic and fermionic modes when working with on-shell-closing multiplets which arise from integrating out auxiliary fields (if they exist). 
\end{itemize}
We further note that adding trivial pairs doesn't affect this correspondence, since they're always one-to-one bosonic--fermionic. This is important at the level of localisation because we require the functional Berezinian associated to the rescaling of normal modes well-behaved in the limit \(t\to\infty\) (compare to \eqref{eq:expansion around localisation locus superparticle} for the superparticle). However, if bosonic and fermionic normal modes don't match, and we apply a naïve uniform rescaling among these modes, the Berezinian will generically be some expression which is either divergent or convergent to zero as \(t\to\infty\).

There appear to be two approaches one could take. The first most often appears in the literature, while the second is new to this paper (to the best of the authors' knowledge):
\begin{itemize}
	\item In the first approach (which we call the \emph{mixed} approach, in contradistinction to the \emph{democratic} approach below), most common in the literature, one starts of by integrating out the Nakanishi--Lautrup field \(b\in\Gamma(\ad P)\) in a \(\delta\)-function gauge, \emph{before} performing any rescalings of normal modes and taking the limit \(t\to\infty\). This eliminates the pure gauge modes, leaving one with at most the discrepancy in bosonic and fermionic modes which arises from the non-closure of the supersymmetry algebra away from the critical locus. This discrepancy can be dealt with through the method introduced in section \ref{ssec:Nilpotent localisation scheme for the on-shell superparticle}, namely a non-uniformly weighted rescaling of the normal modes, dictated by the higher order antifield terms in the BV action.
	\item In the second approach, which we call the \emph{democratic} approach, we treat all fields, including ghosts and trivial pairs, on the same footing. In this approach, we do not integrate out any fields before rescaling the normal modes and taking the limit \(t\to\infty\).	
	Rather, we regard ghost fields as the natural superpartners to pure gauge modes, and we regard trivial pairs as supersymmetry doublets. This also resolves the discrepancy in bosonic and fermionic modes arising from the gauge structure of the theory, again leaving one only with the discrepancy arising from the non-closure of the supersymmetry algebra away from the critical locus. Again, this discrepancy is dealt with using the method introduced in section \ref{ssec:Nilpotent localisation scheme for the on-shell superparticle}.
\end{itemize}
We refer the reader to section \ref{sssec:Localisation argument} for a discussion of this in the case of an off-shell supersymmetry algebra. 
For the reasons described in sections \ref{ssec:BRST reformulation of the off-shell localisation argument} and \ref{sec:superparticle}, we will throughout the rest of this section assume that the space of functionals is ring-theoretically localised to include a multiplicative inverse \(\frac1\varepsilon\) of the global ghost \(\varepsilon\).

\subsubsection{Mixed approach}

In the mixed approach, the combined fermion \(\Psi(t)\) is taken to be of the form
\begin{equation}
	\Psi(t) = \Psi_\gf + \tfrac1\varepsilon\Psi_\loc(t)\,,
\end{equation}
where \(\Psi_\gf\) is the gauge-fixing fermion \eqref{eq:gf fermion Lorenz gauge} imposing the Lorenz gauge as a \(\delta\)-function gauge. Constructing a viable localising fermion for the on-shell multiplet will be slightly more involved, because of the \(O((\hat\lambda^+)^2)\) contributions to the global--gauge BV action \eqref{eq:global-gauge BV action SYM}.

To this end, we perform an isometric bundle decomposition outlined in appendix \ref{Appendix}, which reads
\begin{subequations}
\begin{align}
	\Gamma(S)\phantom{^\ast} &= \Omega_\kappa^1(M) \oplus \Omega_\tH^{(0,1)}(M)\,,
	&
	\zeta &\mapsto \zeta_0\kappa + \zeta_-\bar e\,,
	\\
	\Gamma(S^\ast) &= \Omega_\kappa^1(M) \oplus \Omega_\tH^{(1,0)}(M)\,,
	&
	\hat\zeta &\mapsto \hat\zeta_0\kappa + \zeta_+e\,.
\end{align}
\end{subequations}
Here, \(e\), \(\bar e\) are the complex einbein and its conjugate on the underlying Riemann surface \(\Sigma\), lifted to the Seifert manifold \(M\). The formul\ae{} for these objects are summarised in \eqref{Appendix: spinor bundle isomorphism summary}. Using this decomposition we can rewrite the global--gauge BV action as
\begin{equation}
	S_\gBV = S_\BV + \varepsilon\cQ^\# + \varepsilon^2\int_M\frac{\extd^3x}{\sqrt{\det g}} ~ \tfrac12 g_\ym^2\big[(\hat\lambda_0)^+\big]^2\,.
\end{equation}
A viable choice of localising fermion then reads
\begin{equation}
\label{eq:localising fermion SYM theory}
	g_\ym^2\Psi_\loc(t)
	=
	\int_M t\star \hat\lambda_0(\cQ\hat\lambda)_0 + \big(\tfrac12 t^2 + t\big)\star\hat\lambda_+(\cQ\hat\lambda)^\dagger_-\,.
\end{equation}
The corresponding global--gauge BRST action \(S_\gBRST\) is given by
\begin{equation}
\label{eq:deformed action mixed approach}
\begin{aligned}
	S_\gBRST(t)
	=
	\frac1{g_\ym^2}\int_M
	\begin{aligned}[t]
	&\tfrac12(1+t)^2F_A\wedge\star F_A + \tfrac12(1+t)^2\extD_A\sigma\wedge\star\extD_A\sigma
	\Big.\\
	&+ \star(\zetad\zeta)(1+t)\phantom{^2}\hat\lambda_0\,\big(\rmi\extDs_A - \ad\sigma - \tfrac12 H\big)\lambda_0
	\Big.\\
	&+ \star(\zetad\zeta)(1+t)^2\hat\lambda_+\big(\rmi\extDs_A - \ad\sigma - \tfrac12 H\big)\lambda_-
	\Big.\\
	&+ \star \rmi b(\nabla\inp A) - \nabla\bar c\wedge\star\extD_A c\,.
	\Big.
	\end{aligned}
\end{aligned}
\end{equation}
In this expression, we absorbed the mixing terms \(\varepsilon\cQ\Psi_\gf\) into a field-dependent shift of the ghosts (compare to section \ref{sssec:loc-gf-BRST}). This shift is given by \(c \to c + (\nabla\inp\extD_A)^{-1}(\nabla\inp\varepsilon\cQ A)\), where \((\nabla\inp\extD_A)^{-1}\) denotes appropriate convolution with the Green's function of \(\nabla\inp\extD_A\). This shift evidently induces a unit Jacobian factor in the path integral.

This is a viable choice of localising fermion because the rescaling factors appearing in the global--gauge BRST action allow for a sensible weighted expansion around the localisation locus, allowing one to sensibly take the limit \(t\to\infty\). In the mixed approach, one proceeds to integrate out the Nakanishi--Lautrup field in the path integral expression for a BPS observable,
\begin{equation}
\label{eq:computation localisation mixed case step}
\begin{split}
	\EV{\cO_\BPS}(t)
	&= 
	\sum_{[P]}\int_{\frF_{P|\BRST}}\extD A\,\extD\sigma\,\extD\lambda\,\extD\hat\lambda\,\extD c\,\extD\bar c\,\extD b ~ \cO_\BPS\rme^{-S_\gBRST(t)}
	\\
	&=
	\sum_{[P]}\int_{\frF_{P|\BRST}}\extD A\,\extD\sigma\,\extD\lambda\,\extD\hat\lambda\,\extD c\,\extD\bar c ~ \delta_{\frg_P}\big[\nabla\inp A\big]\cO_\BPS\rme^{-S_\gBRST(t)}
	\\
	&= 
	\sum_{[P]}\int_{\nabla\inp A = 0}\extD A_\gf\,\extD\sigma\,\extD\lambda\,\extD\hat\lambda\,\extD c\,\extD\bar c ~ \cO_\BPS\rme^{-S_\gBRST(t)}\,,
\end{split}
\end{equation}
where \(\delta_{\frg_P}[-]\) the functional \(\delta\)-distribution defined with respect to the natural measure \(\extD\mu_{\frg_P}\) on the quadratic Lie algebra \(\frg_P = \Gamma(\ad P)\), induced by the bilinear form \(\int_M\star\EV{-,-}_\frg\). \(\extD A_\gf = \extD A/\extD\mu_{\frg_P}\) is the measure induced through \(\extD A|_{\nabla\inp A = 0} = \extD A_\gf \otimes \extD\mu_{\frg_P}\) on the gauge-fixed surface \(\cA(P)_\gf = \{A\in\cA(P)\,|\,\nabla\inp A = 0\}\). 

Taking the limit \(t\to\infty\) of action \eqref{eq:deformed action mixed approach}, it is clear that the localisation locus becomes
\begin{align}
\label{eq:localisation locus mixed case}
	F_A &= 0\,,
	&
	\extD_A\sigma &= 0\,.
	&
	\lambda &= \hat\lambda = 0\,,
\end{align}
where for simplicity, we assume that the localisation locus \(\frF_{P|\loc}\) is purely bosonic.
We now expand the normal modes to the localisation locus as
\begin{subequations}
\label{eq:expansion about locus mixed case}
\begin{align}
&
\begin{aligned}
	A_\gf &= A_\circ 
	+ \tfrac1{1+t}\bar\delta A_\gf\,,
	\\
	\sigma &= \sigma_\circ 
	+ \tfrac1{1+t}\bar\delta\sigma\,,
\end{aligned}
&
&
\begin{aligned}
	\lambda_0 &= 0 
	+ \bar\delta\lambda_0\,,
	\\
	\lambda_- &= 0
	+ \bar\delta\lambda_-\,,
\end{aligned}
&
&
\begin{aligned}
	\hat\lambda_0 &= 0
	+ \tfrac1{1+t}\bar\delta\hat\lambda_0\,,
	\\
	\hat\lambda_+ &= 0
	+ \tfrac1{(1+t)^2}\bar\delta\hat\lambda_+\,.
\end{aligned}
\end{align}

After gauge-fixing, \(A_\gf\) carries \(2\times\dim\frg\) bosonic local degrees of freedom and \(\sigma\) carries \(1\times\dim\frg\) bosonic local degrees of freedom. On the other hand, \(\lambda\) and \(\hat\lambda\) each carry \(2\times\dim\frg\) fermionic local degrees of freedom. It follows that the Jacobian factor from rescaling the fluctuations is given by
\begin{equation}
	\Ber\fder{(\delta A_\gf,\delta\sigma,\delta\lambda_0,\delta\lambda_-,\delta\hat\lambda_0,\delta\hat\lambda_+)}{(\bar\delta A_\gf,\bar\delta\sigma,\bar\delta\lambda_0,\bar\delta\lambda_-,\bar\delta\hat\lambda_0,\bar\delta\hat\lambda_+)}
	=
	\Det {(1+t)^{(2 + 1 - 0 - 0 - 1 - 2)\times\dim\frg}} = 1\,,
\end{equation}
where the functional Berezinian is defined by applying \(\zeta\)-function regularisation to harmonic analysis on the underlying spacetime.
Furthermore, the action \eqref{eq:deformed action mixed approach} expands into
\begin{equation}
	S_\gBRST(t) 
	= 
	S_\lin + O\big(\tfrac1{1+t}\big)\,,
\end{equation}
where the linearised action is given by
\begin{equation}
\begin{aligned}
S_\lin 
&= \frac1{g_\ym^2}\int_M
\begin{aligned}[t]
	&\phantom{{}+{}}\tfrac12 \extD_{A_\circ}\bar\delta A_\gf\wedge\star\extD_{A_\circ}\bar\delta A_\gf
	\\
	&+ \tfrac12 \big(\extD_{A_\circ}\bar\delta\sigma + [\bar\delta A_\gf,\sigma_\circ]\big)\wedge\star\big(\extD_{A_\circ}\bar\delta\sigma + [\bar\delta A_\gf,\sigma_\circ]\big)
	\\
	&+ \bar\delta\hat\lambda\big(\rmi\extDs_{A_\circ} - \ad\sigma_\circ - \tfrac12 H\big)\bar\delta\lambda
	\\
	&- \nabla\bar c\wedge\star\extD_{A_\circ}c\,.
\end{aligned}
\end{aligned}
\end{equation}
\end{subequations}
Note that this action is linear over the localisation locus \(\frF_{P|\loc}\), on which the localised fields \((A_\circ,\sigma_\circ)\) are regarded as background fields. From this, we can perform the localisation,
\begin{equation}
\begin{split}
	&\EV{\cO_\BPS}
	= \lim_{t\to\infty}\EV{\cO_\BPS}(t)
	\\
	&\overeq{\ref{eq:computation localisation mixed case step}}
	\lim_{t\to\infty}\sum_{[P]}\int_{\nabla\inp A = 0}\extD A_\gf\,\extD\sigma\,\extD\lambda\,\extD\hat\lambda\,\extD c\,\extD\bar c ~ \cO_\BPS\rme^{-S_\gBRST(t)}
	\\
	&\overeq{\ref{eq:localisation locus mixed case}}
	\lim_{t\to\infty}\sum_{[P]}\int_{\tN\frF_{P|\loc}}
	\big(\extD A_\circ\,\extD\sigma_\circ\big)\big(\extD\delta A_\gf\,\extD\delta\sigma\,\extD\delta\lambda\,\extD\delta\hat\lambda\,\extD c\,\extD\bar c\big) ~ \cO_\BPS\rme^{-S_\gBRST(t)}
	\\
	&\overeq{\ref{eq:expansion about locus mixed case}}
	\sum_{[P]}\int_{\tN\frF_{P|\loc}}
	\big(\extD A_\circ\,\extD\sigma_\circ\big)
	\big(\extD\bar\delta A_\gf\,\extD\bar\delta\sigma\,\extD\bar\delta\lambda\,\extD\bar\delta\hat\lambda\,\extD c\,\extD\bar c\big) ~ \cO_\BPS\rme^{-S_\lin}\,.
\end{split}
\end{equation}
Integrating out the normal fibres, we conclude that the localisation of the BPS observable \(\cO_\BPS\) is given by
\begin{equation}
\label{eq:localised BPS observable mixed case}
	\EV{\cO_\BPS}
	=
	\sum_{[P]}\int_{\frF_{P|\loc}}\extD A_\circ\,\extD\sigma_\circ ~ \frac{\cO_\BPS}{\Ber^{\frac12}\Hess_\perp S_\lin}\,,
\end{equation}
where we defined \(\Hess_\perp\) to be the Hessian along the normal fibres to the localisation locus \(\frF_{P|\loc}\). From here onward, the analysis follows standard sources in the literature \cite{Kapustin:2009kz,Willett:2016adv}.

\subsubsection{Democratic approach}

Next up is the democratic approach. In this approach, we treat ghosts and trivial pairs on the same footing as the fields of the supermultiplet. Accordingly, we take the combined fermion to be
\begin{equation}
	\Psi(t) = (1+t)^2\Psi_\gf + \tfrac1\varepsilon\Psi_\loc(t)\,,
\end{equation}
where \(\Psi_\gf\) is the gauge-fixing fermion \eqref{eq:gf fermion Lorenz gauge} enforcing the Lorenz gauge \(\nabla\inp A = 0\), and \(\Psi_\loc(t)\) the localising fermion constructed in \eqref{eq:localising fermion SYM theory}. Accordingly, the global--gauge BRST action is now given by
\begin{equation}
\label{eq:deformed action democratic approach}
\begin{aligned}
	S_\gBRST(t)
	=
	\frac1{g_\ym^2}\int_M
	\begin{aligned}[t]
	&\tfrac12(1+t)^2F_A\wedge\star F_A + \tfrac12(1+t)^2\extD_A\sigma\wedge\star\extD_A\sigma
	\Big.\\
	&+ \star(\zetad\zeta)(1+t)\phantom{^2}\hat\lambda_0\,\big(\rmi\extDs_A - \ad\sigma - \tfrac12 H\big)\lambda_0
	\Big.\\
	&+ \star(\zetad\zeta)(1+t)^2\hat\lambda_+\big(\rmi\extDs_A - \ad\sigma - \tfrac12 H\big)\lambda_-
	\Big.\\
	&+ \star \rmi(1+t)^2 b(\nabla\inp A) - (1+t)^2\nabla\bar c\wedge\star\extD_A c\,,
	\Big.
	\end{aligned}
\end{aligned}
\end{equation}
where the gauge-fixing terms now also acquire a rescaling factor. Now, rather than first integrating out the Nakanishi--Lautrup field \(b\), we immediately take the limit \(t\to\infty\). 
The bosonic gauge-fixing term, rather than localising due to being positive-semidefinite, localises by rapid oscillations. 
We then find that the theory localises to
\begin{align}
\label{eq:localisation locus democratic case}
	F_A &= 0\,,
	&
	\extD_A\sigma &= 0\,,
	&
	\nabla\inp A &= 0\,,
	&
	b &= 0\,,
	&
	\lambda &= \hat\lambda = c = \bar c = 0\,,
\end{align}
where again for simplicity assume that the localisation locus \(\frF_{P|\loc}\) is bosonic. 
Expansion about the localisation locus is now given by
\begin{subequations}
\label{eq:expansion about locus democratic case}
\begin{equation}
\begin{aligned}
	A &= A_\circ + \tfrac1{1+t}\bar\delta A\,,
	&
	\lambda_0 &= 0 + \bar\delta\lambda_0\,,
	&
	c &= 0 + \tfrac1{1+t}\bar\delta c\,,
	\\
	\sigma &= \sigma_\circ + \tfrac1{1+t}\bar\delta \sigma\,,
	&
	\lambda_- &= 0 + \bar\delta\lambda_-\,,
	&
	\bar c &= 0 + \tfrac1{1+t}\bar\delta\bar c\,,
	\\
	&
	&
	\hat\lambda_0 &= 0 + \tfrac1{1+t}\bar\delta\hat\lambda_0\,,
	&
	b &= 0 + \tfrac1{1+t}\bar\delta b\,.
	\\
	&
	&
	\hat\lambda_+ &= 0 + \tfrac1{(1+t)^2}\bar\delta\hat\lambda_+\,,
\end{aligned}
\end{equation}
Note now, that opposed to the mixed approach, the connection now carries \(3\times\dim\frg\) local degrees of freedom, since it isn't gauge-fixed.
This induces a Jacobian factor
\begin{equation}
	\Ber\fder{(\delta A,\delta\sigma,\delta\lambda_0,\delta\lambda_-,\delta\hat\lambda_0,\delta\hat\lambda_+,\delta c,\delta\bar c,\delta b)}{(\bar\delta A,\bar\delta\sigma,\bar\delta\lambda_0,\bar\delta\lambda_-,\bar\delta\hat\lambda_0,\bar\delta\hat\lambda_+,\bar\delta c,\bar\delta\bar c,\bar\delta b)}
	=
	\Det{(1+t)^{(3 + 1 - 0 - 0 - 1 - 2 - 1 - 1 + 1)\times \dim\frg}}
	=
	1\,.
\end{equation}
Now, the triviality of this Jacobian factor follows from the rescaling of the ghosts countering the rescaling of the aditional pure gauge modes of the gauge potential, alongside the rescalings of the trivial pairs cancelling each other out. Expanding the global--gauge BV action \eqref{eq:deformed action democratic approach} now yields
\begin{equation}
	S_\gBRST(t)
	=
	S_\lin + O\big(\tfrac1{1+t}\big)\,,
\end{equation}
where the linearised action is now given by
\begin{equation}
\begin{aligned}
S_\lin 
&= \frac1{g_\ym^2}\int_M
\begin{aligned}[t]
	&\phantom{{}+{}}\tfrac12 \extD_{A_\circ}\bar\delta A\wedge\star\extD_{A_\circ}\bar\delta A
	\\
	&+ \tfrac12 \big(\extD_{A_\circ}\bar\delta\sigma + [\bar\delta A,\sigma_\circ]\big)\wedge\star\big(\extD_{A_\circ}\bar\delta\sigma + [\bar\delta A,\sigma_\circ]\big)
	\\
	&+ \bar\delta\hat\lambda\big(\rmi\extDs_{A_\circ} - \ad\sigma_\circ - \tfrac12 H\big)\bar\delta\lambda
	\\
	&+ \star\rmi\bar\delta b(\nabla\inp\bar\delta A) - \nabla\bar c\wedge\star\extD_{A_\circ}c\,.
\end{aligned}
\end{aligned}
\end{equation}
\end{subequations}
Similarly, we now perform the localisation as
\begin{equation}
\begin{split}
	&\EV{\cO_\BPS}
	= \lim_{t\to\infty}\EV{\cO_\BPS}(t)
	\\
	&= \lim_{t\to\infty}\sum_{[P]}\int_{\frF_{P|\BRST}}
	\extD A\,\extD\sigma\,\extD\lambda\,\extD\hat\lambda\,\extD c\,\extD\bar c\,\extD b ~ \cO_\BPS\rme^{-S_\gBRST(t)}
	\\
	&\overeq{\ref{eq:localisation locus democratic case}} \lim_{t\to\infty}\sum_{[P]}\int_{\tN\frF_{P|\loc}}\big(\extD A_\circ\,\extD\sigma_\circ\big)\big(\extD\delta A\,\extD\delta\sigma\,\extD\delta\lambda\,\extD\delta\hat\lambda\,\extD\delta c\,\extD\delta\bar c\,\extD\delta b\big) ~ \cO_\BPS\rme^{-S_\gBRST(t)}
	\\
	&\overeq{\ref{eq:expansion about locus democratic case}}
	\sum_{[P]}\int_{\tN\frF_{P|\loc}}\big(\extD A_\circ\,\extD\sigma_\circ\big)\big(\extD\bar\delta A\,\extD\bar\delta\sigma\,\extD\bar\delta\lambda\,\extD\bar\delta\hat\lambda\,\extD\bar\delta c\,\extD\bar\delta\bar c\,\extD\bar\delta b\big) ~ \cO_\BPS\rme^{-S_\lin}
\end{split}
\end{equation}
Integrating out the normal fibres, one now finds that the localisation of the expectation value of the BPS observable is given by
\begin{equation}
	\EV{\cO_\BPS}
	=
	\sum_{[P]}\int_{\frF_{P|\loc}}\extD A_\circ\,\extD\sigma_\circ ~ \frac{\cO_\BPS}{\Ber^{\frac12}\Hess_\perp S_\lin}\,.
\end{equation}
While this expression looks identical to \eqref{eq:localised BPS observable mixed case}, and does in fact give the same result, a key difference is now that the Hessian of the linearised action includes the term which imposes the gauge condition.

\subsubsection{Localisation as an \(R_\xi\)-gauge}
\label{ssec:R xi gauge localisation SYM}

Finally, we describe how to realise the localisation computation for a BPS observable \(\cO_\BPS\) as an \(R_\xi\)-gauge (compare to section \ref{ssec:R xi gauge localisation superparticle}). To this end, we introduce two trivial pairs \((\bar\zeta,\beta) \in \tT[1]\Gamma(\Pi S\otimes \ad P)[-1]\) and \((\hat{\bar\zeta},\hat\beta) \in \tT[1]\Gamma(S^\ast\otimes \ad P)[-1]\) of respective bidegrees \(|\bar\zeta| = (-1,\odd)\), \(|\beta| = (0,\odd)\), \(|\hat{\bar\zeta}| = (-1,\even)\) and \(|\hat\beta| = (0,\even)\). Accordingly, we extend the global--gauge BV action \eqref{eq:global-gauge BV action SYM} to
\begin{equation}
\begin{aligned}
	S_\gBV|_\old \to 
	\begin{aligned}[t]
	&S_\gBV|_\new 
	= S_\gBV|_\old + \int_M\extd^3x ~ \big(\rmi\hat\beta\hat{\bar\zeta}^+ - \bar\zeta^+\beta\big)
	\\
	&= S_\gBV|_\old + \int_M\extd^3x ~ \Big[
	\rmi\hat\beta_0(\hat{\bar\zeta}_0)^+ 
	+ \rmi\hat\beta_+(\hat{\bar\zeta}_+)^+ 
	+ \beta_0(\bar\zeta_0)^+
	+ \beta_-(\bar\zeta_-)^+
	\Big]\,.
	\end{aligned}
\end{aligned}
\end{equation}
Following the game plan from section \ref{ssec:R xi gauge localisation superparticle}, a viable choice of combined fermion then reads
\begin{equation}
	g_\ym^2\Psi(t) =
	\int_M\star\Big[
	\bar c(\nabla\inp A)
	-\varepsilon\hat{\bar\zeta}\bar\zeta 
	+ \hat{\bar\zeta}(\cQ\hat\lambda)^\dagger
	+ t \cdot \hat\lambda_0\bar\zeta_0
	+ \big(\tfrac12t^2 + t\big)\cdot \hat\lambda_+\bar\zeta_-
	\Big]\,.
\end{equation}
This combined fermion gives rise to a global--gauge BRST action
\begin{equation}
\begin{aligned}
	S_\gBRST(t)
	&=
	S
	+
	\frac1{g_\ym^2}\int_M\star\Big[
	\begin{aligned}[t]
	&\rmi b(\nabla\inp A) + \bar c(\nabla\inp\extD_A)c
	\Big.\\
	&+ \rmi\hat\beta_0(\zetad\zeta)\Big((\cQ\hat\lambda)_0 - {\varepsilon\zetab_0}\Big) 
	+ \rmi\hat\beta_+(\zetad\zeta)\Big((\cQ\hat\lambda)_-^\dagger - {\varepsilon\zetab_-}\Big)
	\Big.\\
	&+ \beta_0\Big(t\hat\lambda_0 - (\zetad\zeta)({\varepsilon\hat\zetab_0})\Big)
	+ \beta_-\Big(\big(\tfrac12t^2+t\big)\hat\lambda_+ - (\zetad\zeta)({\varepsilon\hat\zetab_+})\Big)
	\Big.\\
	&- (\zetad\zeta)({\varepsilon\hat\zetab_0})\cQ(\cQ\hat\lambda)_0 
	- (\zetad\zeta)({\varepsilon\hat\zetab_+})\cQ(\cQ\hat\lambda)^\dagger_-
	\Big.\\
	&+ t(\cQ\hat\lambda)_0({\varepsilon\zetab_0}) 
	+ \tfrac12t^2({\varepsilon\zetab_0})^2 
	+ \big(\tfrac12 t^2 + t\big)(\cQ\hat\lambda)_+({\varepsilon\zetab_-})
	\Big]\,.
	\end{aligned}
\end{aligned}
\end{equation}
Integrating out the Nakanishi--Lautrup fields \(\beta_0,\beta_-,\hat\beta_0,\hat\beta_+\) we recover the global--gauge BRST action \eqref{eq:deformed action mixed approach} for the mixed approach\footnote{+Strictly speaking, to integrate out the fields correctly, one has to work in complexified field space and integrate along an appropriate countour correlating \(\hat\beta_+\) and \(\varepsilon\bar\zeta_-\) such that the integration is well-defined. We will not further elaborate on these subtleties. An explicit example of this can be found in \cite[\S{}3]{Pestun:2007rz}.}.
At \(t = 0\), this confirms physical equivalence with the original theory, and at \(t > 0\) this is manifestly equivalent to the previous approaches.
To realise the localisation as a \(\delta\)-function gauge, we perform rescalings
\begin{align}
	\bar\zeta_0 &\mapsto \tfrac1{t}\bar\zeta_0\,,
	&
	\beta_0 &\mapsto \tfrac1{t}\beta_0\,,
	&
	\bar\zeta_- &\mapsto \tfrac2{t^2 + 2t}\bar\zeta_-\,,
	&
	\beta_- &\mapsto \tfrac2{t^2 + 2t}\beta_-\,,
\end{align}
which clearly has a unit functional Jacobian factor associated to it. Taking the limit \(t\to\infty\), the global--gauge BRST action becomes
\begin{equation}
\begin{aligned}
	S_\gBRST(t)
	\xrightarrow{t\to\infty}
	S
	+
	\frac1{g_\ym^2}\int_M\star\Big[
	\begin{aligned}[t]
	&\rmi b(\nabla\inp A) + \bar c(\nabla\inp\extD_A)c
	\Big.\\
	&+ \rmi\hat\beta(\cQ\hat\lambda)^\dagger
	- \frac1{\zetad\zeta}\hat\lambda\beta
	- (\varepsilon\hat\zetab)\cQ(\cQ\hat\lambda)^\dagger
	\Big.\\
	&+ (\cQ\hat\lambda)_0({\varepsilon\zetab_0}) + \tfrac12({\varepsilon\zetab_0})^2 + (\cQ\hat\lambda)_+({\varepsilon\zetab_-})
	\Big]\,.
	\end{aligned}
\end{aligned}
\end{equation}
In this limit, all of the components of the trivial pairs become Lagrange multipliers. Again, to correctly integrate these out we need to integrate over an appropriate contour in complexified field space. Integrating out the trivial pairs \((\bar c,b)\), \((\bar\zeta,\beta)\) and \((\hat{\bar\zeta},\hat\beta)\), we then find that the expectation value of the BPS observable \(\cO_\BPS\) localises to
\begin{equation}
\begin{split}
	\EV{\cO_\BPS}
	&=
	\lim_{t\to\infty}\sum_{[P]}\int_{\frF_{P|\BRST}}\extD A\,\extD\sigma\,\extD\lambda\,\extD\hat\lambda\,\extD c\,\extD\bar c\,\extD b\,\extD(\varepsilon\bar\zeta)\,\extD\beta\,\extD(\varepsilon\hat{\bar\zeta})\,\extD\hat\beta ~ \cO_\BPS\rme^{-S_\gBRST(t)}
	\\
	&= \sum_{[P]}\int_{\frF_{P|\BRST}}\extD A\,\extD\sigma\,\extD\lambda\,\extD\hat\lambda\,\extD c ~ \cO_\BPS
	\begin{aligned}[t]
	&\times \delta\big[\nabla\inp A\big]\delta\big[(\nabla\inp\extD_A)c\big]
	\\
	&\times \delta\big[(\cQ\hat\lambda)_0\big]
	\delta\big[(\cQ\hat\lambda)^\dagger_-\big]
	\delta\big[(\cQ\hat\lambda)_+\big]
	\\
	&\times \delta[\hat\lambda_0]\delta[\hat\lambda_+]\delta\big[\cQ(\cQ\hat\lambda)_0\big]\delta\big[\cQ(\cQ\hat\lambda)^\dagger_-\big]
	\end{aligned}
\end{split}
\end{equation}
We note that, by using the properties \eqref{eq:properties delta functions} of the Dirac \(\delta\)-functionals, we arrive at the same result as in the other sections. Thus, we conclude that localisation can be implemented as a \(\delta\)-function gauge for on-shell-closing supersymmetry algebras in a gauge theory. The use of trivial pairs in the equivariant rather than nilpotent case will be treated in future work.

\section{Conclusions}
The picture  of localisation and gauge fixing within the Batalin--Vilkovisky formalism developed here provides a unified conceptual framework  that is  computationally convenient in various regards. In particular, by placing global and gauge symmetries on an equal homological footing, the formalism naturally accommodates on-shell realisations of supersymmetry. We have presented explicit examples illustrating this point for the Witten index of the \(d=1\) superparticle and the partition function of \(d=3\), \(\mathcal{N}=2\) supersymmetric Yang--Mills theory. This sets up a number of immediate generalisations and applications that will be treated in future work. For a summary of these results and future directions, see \cref{sec:intro}.

Here, instead, we conclude by briefly mentioning  some more speculative and ambitious possibilities, extending beyond rigid supersymmetric field theories.  In previous work \cite{Borsten:2025hrn}, we emphasised the common structures underpinning supersymmetric twists, spontaneous symmetry breaking, anomalies, and localisation, all as instances of  twisted $L_\infty$-algebras.  Combined with the Batalin--Vilkovisky formalism for localisation developed here, this suggests natural applications to localisation in twisted supergravity \cite{Costello:2016mgj} and the localisation of supergravity in certain classical backgrounds (both of which  may be regarded as global versions of $L_\infty$-algebra twists). In the former case, twisting corresponds to giving the local supersymmetry ghost a vacuum expectation value, which is conjectured to provide a unique quantisation in perturbation theory for  type II supergravity, building on the results of \cite{Costello:2015xsa}. Our approach may provide novel methods to probe this statement.   In the latter case, perhaps most enticing are the recent applications  of supersymmetric localisation to the quantum entropy of  black holes in supergravity \cite{Dabholkar:2010uh,Dabholkar:2011ec,Jeon:2018kec,Cassani:2025sim}. In these works, off-shell formulations of the superalgebras are required. While there are such formulations for certain models, it would be desirable to go beyond these. Our framework suggests a clear, albeit technically nuanced, path.

\appendix

\section{Spinor calculus on Seifert manifolds}
\label{Appendix}

\paragraph{Spinors.}
In \(d=3\), the gamma matrices \(\gamma^a\) (\(a = 1,2,3\)) are the Pauli matrices.
Spinors correspond to vectors \(\zeta\) and their conjugates \(\hat\zeta\) in the fundamental representation of \(SU(2)\).
The charge conjugation matrix is \(\sfC=\mathrm i\gamma^2=
	(\begin{smallmatrix}
		0 & 1
		\\
		-1 & 0
	\end{smallmatrix})\), using which the charge conjugation of spinors is defined as
\begin{equation}\begin{aligned}
		\zeta^\sfc &= (\sfC\zeta)^\sft\,,&
		\hat\zeta^\sfc &= (\hat\zeta\sfC)^\sft\,.
	\end{aligned}
\end{equation}
For a \(2\times2\) matrix \(M\), the Fierz identity reads
\begin{equation}
\label{Fierz identity 3d}
	M = \tfrac12\tr(M)\idop + \tfrac12\tr(M\gamma_a)\gamma^a\,.
\end{equation}
Applied to the matrix \(\zeta\otimes\zetat\), this yields
\begin{align}
	\zeta\otimes\hat\zeta
	&=
	\tfrac12 \hat\zeta\zeta(\idop + \Ks)\,,
	&
	K^a &= \frac{1}{\hat\zeta\zeta}\hat\zeta\gamma^a\zeta\,.
\end{align}

\paragraph{Complex structure.} The contact structure \(\kappa\) \eqref{eq:seifert contact structure} defines a connection on the Seifert manifold \(M\to \Sigma\). The tangent space then decomposes into vertical and horizontal subspaces as \(\tT M = \mathrm VM \oplus \tH M\). At the level of 1-forms we may decompose
\begin{align}
	\Omega^1(M,\bbC) &= \Omega^1_\kappa(M) \oplus \Omega^1_\tH(M)\,,
	&
	\begin{aligned}
		\Omega^1_\kappa(M) &= \ker(\sfi\circ\sfp - 1)
		\\
		\Omega^1_\tH(M) &= \ker\sfp
	\end{aligned}
\end{align}
where we have defined inclusion and projection operators
\begin{subequations}
\begin{align}
	\sfi &: \Omega^0(M,\bbC) \into \Omega^1(M,\bbC)\,,
	&
	\alpha_0 &\mapsto \kappa\alpha_0
	\\
	\sfp &: \Omega^1(M,\bbC) \onto \Omega^0(M,\bbC)\,,
	&
	\alpha_1 &\mapsto K\inp\alpha_1
\end{align}
\end{subequations}
allowing us to write 1-forms \(\alpha_1\in\Omega^1(M)\) as 
\begin{align}
	\alpha_1 &= \kappa\alpha_0 + \alpha_\tH\,,
	&
	&
	\begin{aligned}
		&\alpha_0 \in C^\infty(M,\bbC)
		\\
		&K\inp\alpha_\tH = 0
	\end{aligned}\,.
\end{align}
We can now define a complex structure on \(\Omega^1_\tH(M)\) through the operator
\begin{align}
	J &= K\inp\star : \Omega^1_\tH(M) \to \Omega^1_\tH(M)\,,
	&
	J^2 &= -1\,.
\end{align}
We define the projection operators corresponding to the \(\pm\mathrm i\) eigenspaces of \(J\) on \(\Omega^1_\tH(M)\)
\begin{equation}
	\cP^J_\pm \coloneq \frac{1 \mp\mathrm iJ}{2} : \Omega^1_\tH(M) \to \Omega^1_\tH(M)\,,
\end{equation}
allowing us to decompose further as
\begin{equation}
\label{decomposition 1-forms Seifert manifold}
	\Omega^1(M,\bbC) = \Omega^1_\kappa(M) \oplus \Omega^{(1,0)}_\tH(M) \oplus \Omega^{(0,1)}_\tH(M)\,,
\end{equation}
where
\begin{align}
	\Omega_\tH^{(1,0)}(M) &\coloneqq \Im\cP_+^J\,,&
	\Omega_\tH^{(0,1)}(M) &\coloneqq \Im\cP_-^J\,.
\end{align}
As \(C^\infty(M,\bbC)\)-modules, the spaces \(\Omega_\tH^{(1,0)}(M)\) and \(\Omega_\tH^{(0,1)}(M)\) are one-dimensional and can be taken to be generated by respectively by the nowhere vanishing form \(e = e^1 + \mathrm ie^2\in\Omega^{(1,0)}_\tH(M)\) and its complex conjugate \(\eb = e^1 {-}\mathrm ie^2 \in\Omega^{(0,1)}_\tH(M)\), which together with \(\kappa = e^3\) we will regard as the dreibein of \((M,\star)\). That is, we take these to satisfy the following relations:
\begin{align}
\label{Hodge star dreibeins}
	\star e &= +\mathrm i\kappa\wedge e\,,&
	\star \eb &= -\mathrm i\kappa\wedge \eb\,,&
	\star\kappa &= \tfrac12\mathrm ie\wedge\eb\,,&
	\star 1 &= \tfrac12\mathrm i\kappa\wedge e\wedge\eb\,.
\end{align}
An arbitrary form \(\alpha_1 \in \Omega^1(M,\bbC)\) can then be uniquely decomposed as
\begin{align}
	\alpha_1 &= \kappa\alpha_0 + \tfrac12 e\alpha_+ + \tfrac12 \eb\alpha_-\,,
	&
	\alpha_0,\alpha_\pm &\in C^\infty(M,\bbC)\,.
\end{align}
For real 1-forms, \(\alpha_0\) is real-valued, and \(\alpha_\pm\) are related by complex conjugation.

\paragraph{Decomposing spinors.} In this paragraph we construct the useful isomorphisms
\begin{align}\label{spinor decomposition}
	\Gamma(S) &\cong \Omega^1_\kappa(M) \oplus \Omega^{(0,1)}_\tH(M)\,, &
	\Gamma(S^\star) &\cong \Omega^1_\kappa(M) \oplus \Omega^{(1,0)}_\tH(M)\,.
\end{align}
A convenient choice of a zweibein is
\begin{align}
\label{zweibein expression}
	e &= -\frac{1}{\zetad\zeta}\zetac\gamma\zeta
	\,,&
	\eb &= -\frac{1}{\zetad\zeta}\zetad\gamma\zetacd
	\,.
\end{align}
Then, by repeatedly applying the Fierz identity \eqref{Fierz identity 3d} along with properties of the contact structure, one can obtain the spinor bundle decompositions, the decomposition \eqref{spinor decomposition} is such that, for \(\eta\in\Gamma(S)\) and \(\hat\eta\in\Gamma(S^*)\),
\begin{equation}
\label{Appendix: spinor bundle isomorphism summary}
\begin{aligned}
	\eta &\mapsto \kappa\eta_0 + \eb\eta_-\,,
	&
	\eta_0 &= \frac{\zetad\eta}{\zetad\zeta}\,,
	&
	\eta_- &= \frac{\zetac\eta}{\zetad\zeta}\,,
	\\
	\hat\eta &\mapsto \kappa\hat\eta_0 + e\hat\eta_+\,,
	&
	\hat\eta_0 &= \frac{\phantom{^\dagger}\hat\eta\zeta}{\zetad\zeta}\,,
	&
	\hat\eta_+ &= \frac{\hat\eta\zetacd}{\zetad\zeta\phantom{^\sfc}}\,,
\end{aligned}
\end{equation}
so that
\begin{equation}
	\hat\eta\eta
	=
	(\zetad\zeta)\big(\hat\eta_0\eta_0 + \hat\eta_+\eta_-\big)\,.
\end{equation}

\renewcommand\acknowledgments{\section*{Acknowledgements}}
\acknowledgments

This paper benefited from a collaboration with Alexandros Spyridon Arvanitakis\textsuperscript{\orcidlink{0000-0002-6646-5500}}, whose results are contained in a separate paper \cite{Arvanitakis:2025} submitted to the arXiv at the same time as this one. The authors thank Fridrich Valach\textsuperscript{\orcidlink{0000-0003-0020-1999}}, Luigi Alfonsi\textsuperscript{\orcidlink{0000-0001-5231-2354}}, Charles Strickland-Constable\textsuperscript{\orcidlink{0000-0003-0294-1253}}, Charles Alastair Stephen Young\textsuperscript{\orcidlink{0000-0002-7490-1122}}, and Ingmar Akira Saberi\textsuperscript{\orcidlink{0000-0002-2005-938X}} for helpful conversations. Leron Borsten and Hyungrok Kim thank Christian Saemann\textsuperscript{\orcidlink{0000-0002-5273-3359}} and Martin Wolf\textsuperscript{\orcidlink{0009-0002-8192-3124}} for helpful discussions. Leron Borsten is grateful for the hospitality of the Theoretical Physics group, Blackett Laboratory, Imperial College London.

\bibliographystyle{JHEP}
\bibliography{biblio}

\end{document}